\documentclass[twocolumn]{aastex63}

\newcommand{\oi}{[O\,{\scriptsize I}]}
\newcommand{\neii}{[Ne\,{\scriptsize II}]}
\newcommand{\neiii}{[Ne\,{\scriptsize III}]}
\newcommand{\sii}{[S\,{\scriptsize II}]}
\newcommand{\dem}[1]{\textcolor{gray}{#1}}

\shorttitle{Evolution of Disk Winds}
\shortauthors{Pascucci et al.}

\begin{document}
\title{The Evolution of Disk Winds from a Combined Study of Optical and Infrared Forbidden Lines}

\correspondingauthor{Ilaria Pascucci}
\email{pascucci@lpl.arizona.edu}

\author{Ilaria Pascucci}
\affiliation{Lunar and Planetary Laboratory, The University of Arizona, Tucson, AZ 85721, USA}
\affiliation{Earths in Other Solar Systems Team, NASA Nexus for Exoplanet System Science}

\author{Andrea Banzatti}
\affiliation{Department of Physics, Texas State University, 749 N Comanche Street, San Marcos, TX 78666, USA}

\author{Uma Gorti}
\affiliation{SETI Institute/NASA Ames Research Center, Mail Stop 245-3, Moffett Field, CA 94035-1000, USA}

\author{Min Fang}
\affiliation{California Institute of Technology, Cahill Center for   Astronomy and Astrophysics, MC 249-17, Pasadena, CA 91125}

\author{Klaus Pontoppidan}
\affiliation{Space Telescope Science Institute,  3700 San Martin Drive, Baltimore, MD 21218, USA}

\author{Richard Alexander}
\affiliation{School of Physics and Astronomy, University of Leicester, Leicester, LE1 7RH, UK}

\author{Giulia Ballabio}
\affiliation{School of Physics and Astronomy, University of Leicester, Leicester, LE1 7RH, UK}

\author{Suzan Edwards}
\affiliation{Five College Astronomy Department, Smith College, Northampton, MA 01063, USA}

\author{Colette Salyk}
\affiliation{Department of Physics and Astronomy, Vassar College, 124 Raymond Avenue, Poughkeepsie, NY 12604, USA}

\author{Germano Sacco}
\affiliation{INAF – Osservatorio Astrofisico di Arcetri, Largo E. Fermi, 5, 50125 Firenze, Italy}

\author{Ettore Flaccomio}
\affiliation{INAF – Osservatorio Astronomico di Palermo, Piazza del Parlamento 1, 90134 Palermo, Italy}

\author{Geoffrey A. Blake}
\affiliation{Division of Geological and Planetary Sciences, California Institute of Technology, Pasadena, CA 91125, USA}

\author{Andres Carmona}
\affiliation{Astrophysics \& Planetology Research Institute, 9, avenue du Colonel Roche BP 44346 31028 Toulouse Cedex 4}

\author{Cassandra Hall}
\affiliation{School of Physics and Astronomy, University of Leicester, Leicester, LE1 7RH, UK}
\affiliation{Department of Physics and Astronomy, The University of Georgia, Athens, GA 30602, USA}
\affiliation{Center for Simulational Physics, The University of Georgia, Athens, GA 30602, USA}

\author{Inga Kamp}
\affiliation{Kapteyn Astronomical Institute, University of Groningen, Groningen, The Netherlands}

\author{Hans Ulrich K{\"a}ufl}
\affiliation{European Southern Observatory, Karl-Schwarzschild-Str. 2, 85748 Garching, Germany}

\author{Gwendolyn Meeus}
\affiliation{Department of Theoretical Physics, Autonomous of Madrid, Cantoblanco, 28049 Madrid, Spain}
\affiliation{Centro de Investigaci\'on Avanzada en F\'isica Fundamental (CIAFF), Facultad de Ciencias, UAM, 28049 Madrid, Spain}

\author{Michael Meyer}
\affiliation{Department of Astronomy, University of Michigan, 311 West Hall, 1085 S. University Ave, Ann Arbor, MI 48109, USA}

\author{Tyler Pauly}
\affiliation{Space Telescope Science Institute,  3700 San Martin Drive, Baltimore, MD 21218, USA}

\author{Simon Steendam}
\affiliation{Kapteyn Astronomical Institute, University of Groningen, Groningen, The Netherlands}

\author{Michael Sterzik}
\affiliation{European Southern Observatory, Karl-Schwarzschild-Str. 2, 85748 Garching, Germany}



\begin{abstract}
We analyze high-resolution ($\Delta v \leq 10$\,km/s) optical and infrared spectra covering the \oi{} 6300\,\AA{} and \neii{} 12.81\,\micron{} lines from a sample of 31 disks in different evolutionary stages. Following work at optical wavelengths, we use Gaussian profiles to fit the \neii{} lines and classify  them into HVC (LVC) if the line centroid is more (less) blueshifted than 30\,km/s with respect to the stellar radial velocity.
Unlike for the \oi{} where a HVC is often accompanied by a LVC, all 17 sources with a \neii{} detection have either a HVC {\it or} a LVC. \neii{} HVCs are preferentially detected toward high accretors ($\dot{M}_{\rm acc} > 10^{-8}$\,M$_\odot$/yr) while LVCs are found in sources with low $\dot{M}_{\rm acc}$, low \oi{} luminosity,  and large infrared spectral index ($n_{\rm 13-31}$). Interestingly, the \neii{} and \oi{} LVC luminosities display an opposite behaviour with  $n_{\rm 13-31}$: as the inner dust disk depletes (higher $n_{\rm 13-31}$) the \neii{} luminosity increases while the \oi{} weakens. The \neii{} and \oi{} HVC profiles are generally similar with centroids and FWHMs showing the expected behaviour from shocked gas in micro-jets. In contrast, the \neii{} LVC profiles are typically more blueshifted and narrower than the \oi{} profiles. The FWHM and centroid vs. disk inclination suggest that the \neii{} LVC predominantly traces unbound gas from a slow, wide-angle wind that has not lost  completely the Keplerian signature from its launching region. We sketch an evolutionary scenario that could explain the combined \oi{} and \neii{} results and includes screening of hard ($\sim$\,1\,keV) X-rays in inner, mostly molecular, MHD winds.
\end{abstract}



\section{Introduction} \label{sec:intro}
Young ($\sim 1-10$\,Myr) stars are often surrounded by disks of gas and dust within which planets form, hence the term protoplanetary disks. While planet formation contributes to reduce the primordial disk mass, significant mass is removed via accretion of gas through the disk. The final disk clearing is attributed to high-energy stellar photons driving a thermal, also called photoevaporative, wind beyond  a few au
\citep[e.g.,][for a review]{Alexander2014} but how disk gas sheds angular momentum to accrete onto the central star remains a crucial, yet unanswered, question.

 While MHD disk winds were considered early on \citep[e.g.,][]{PelletierPudritz1992}, the prevailing view has been that turbulence driven by  magnetorotational instability \citep[MRI,][]{BalbusHawley1991} transports angular momentum outward, enabling disk material to flow radially inward  \citep[e.g.,][for a review]{Armitage2011}. However,  more recent disk simulations, which include non-ideal MHD effects, find that MRI is suppressed in most of the planet-forming region ($\sim 1-20$\,au); hence accretion is shut off \citep[e.g.,][for a review]{Turner2014}. Corroborating the theoretical results, recent ALMA observations suggest that the majority of the observed disks are weakly turbulent at tens of au  \cite[e.g.,][]{Teague2016,Flaherty2017,Flaherty2020}.
Interestingly, these  non-ideal MHD simulations persistently predict the presence of disk winds, defined as outflowing gas from a few scale heights above the disk midplane. The simulated winds extract enough angular momentum to drive accretion at the observed rates  \citep[e.g.,][]{Gressel2015,Bai2016,Gressel2020}. These outer winds (beyond a few au out to tens {\bf of} au in some models), combined with the closer-in winds likely responsible for outflowing gas at hundreds of km/s \citep[hereafter jets, e.g.,][]{Frank2014}, could drive disk evolution with important implications for planet formation and migration \citep[e.g.,][]{Ogihara2018,Kimmig2020}.

Identifying disk winds requires finding gas lines that trace the unbound disk surface at a spectral resolution sufficient enough to detect velocity shifts with respect to Keplerian motion around the star \citep[e.g.,][for a recent review on disk winds]{ErcolanoPascucci2017}. The first evidence for slow, likely thermal, winds came from high signal-to-noise, high-resolution ($\Delta v \sim$10\,km/s) spectra of \neii{} at 12.81\,\micron{} in three disks with inner dust cavities \citep{PascucciSterzik2009,Pascucci2011}. The emission lines have modest widths (FWHM$\sim$15-40\,km/s) and small blueshifts ($\sim$3-6\,km/s) in the centroid velocity, compatible with earlier predictions from thermally-driven photoevaporative flows  \citep{Alexander2008}. Another 10 sources show similar profiles \citep{Sacco2012,Baldovin2012} but the lack of sensitive high-resolution mid-infrared spectrographs has so far precluded gathering a diverse and large sample of disks to identify evolutionary trends.

More progress could be made at optical wavelengths. Optical forbidden lines, such as the \oi{} at 6300\,\AA{}, have long been known to possess a so-called low velocity component (LVC), emission  blueshifted by less than $\sim$30\,km/s, in addition to a high-velocity component tracing fast ($\sim$100\,km/s) collimated micro-jets \citep[e.g.,][]{Hartigan1995}. Early on, \citet{KwanTademaru1995} investigated the possibility that the LVC might trace a slow disk wind. More recently, high-resolution ($\Delta v <$10\,km/s) spectroscopy enabled identifying broad wings plus a narrow peak in about half of the LVCs, a profile that has been described as the combination of two Gaussian profiles  \citep[a “broad component,” BC, and a “narrow component,” NC;][]{Rigliaco2013,Simon2016,McGinnis2018}.  \citet{Simon2016} pointed out that the BC, with its large FWHM, cannot trace a thermal wind beyond a few au but, most likely, probes a closer-in MHD wind. \citet{Banzatti2019} further argued that the entire LVC traces a radially extended MHD wind that feeds a jet based on the finding that the kinematic properties of the BC, peak centroid and FWHM, correlate with those of the NC, and the BC and NC kinematics correlate with the equivalent widths of the HVC. However,
\citet{Weber2020} recently suggested that these correlations can be explained by a single common correlation between line luminosity and accretion luminosity, with accretion introduced as an EUV component heating the line emitting region in both their analytic MHD and their X-ray photoevaporative models.
They conclude that optical forbidden line profiles are best reproduced by the combination of an inner MHD wind (producing the HVC and BC) and a photoevaporative wind (producing the NC).

Taking advantage of the recent upgrade of VISIR on the VLT (hereafter VISIR~2), our group carried out a large high-resolution spectroscopic survey of protoplanetary disks (PID: 198.C-0104, PI. K. Pontoppidan) focusing on strong rotational lines of water, H$_2$, and ro-vibrational lines of OH to investigate disk chemistry, and on the \neii{} line at 12.81\,\micron{} to expand the sample of disk wind sources. Here, we focus on the \neii{} observations and connect the outer winds probed by this infrared forbidden line with the winds traced by the \oi{} 6300\,\AA{} transition, the strongest of the optical forbidden lines  \citep[e.g.,][]{Hartigan1995}. First, we describe our combined \oi{} \& \neii{} disk sample (Section~\ref{sec:sample}) and the detections or upper limits from our VISIR~2 survey (Section~\ref{sec:NeII}). Next, we explore if the known correlations between \oi{} luminosities  and stellar/disk properties apply to the \neii{} line luminosities, as well as compare line profiles when both transitions are detected (Section~\ref{sec: results}). Finally, we  discuss our main results which include evidence for evolution in disk winds (Section~\ref{sec:discussion}).

\section{Sample and main properties} \label{sec:sample}
We start from the sample of disks observed with VISIR~2 as part of the Large Program ``Protoplanetary disks as chemical factories'' (PID: 198.C-0104), which includes 40 sources observed in the \neii{} 12.81\,\micron{} setting\footnote{This number includes wide binaries extracted separately.}. Observations were taken with a slit width of $0.75 ''$ delivering a spectral resolution of R$\sim$30,000 ($\sim$10\,km/s) and data were reduced following \citet{Banzatti2014}.  This large program and the data reduction adapted to the new VISIR detector are described in detail in a forthcoming paper (Banzatti et al. in prep).

To the VISIR~2 sample we add all other published VISIR~1 spectra of disks covering the \neii{} line at 12.81\,\micron{}, observed at a similar spectral resolution of 10\,km/s \citep[][]{PascucciSterzik2009, Pascucci2011,Baldovin2012,Sacco2012}.
Next, we cross-match this list with published \oi{} 6300\,\AA{} detections attained at slightly higher spectral resolution  \citep[R$\sim$45,000 or $\sim 7$\,km/s,][]{Simon2016,Fang2018,Banzatti2019} and found 24 common sources. Similar high-resolution optical spectra for an additional 7 disks are retrieved from the archive, their reduction and analysis are described in Appendix~\ref{app:archive}. The resulting sample of 31 disks with \oi{} 6300\,\AA{} detections and VISIR spectra covering the \neii{} 12.81\,\micron{} line is summarized in Table~\ref{tab:targets}. Most of the disks belong to the nearby star-forming regions of Taurus, Lupus, Ophiuchus, Chamaeleon, and Corona Australis.

For each source we collect its 2MASS equatorial position-based name and used it to retrieve the {\it Gaia} Data Release 2 (hereafter, GDR2) parallactic distance from the geometric-distance table generated by \cite{bailer18}.
No parallax is reported for VW~Cha and T~CrA, hence we take as distance that of the star-forming regions the sources belong to: Chamaeleon~I (190\,pc, Roccataglia et al. 2018) and Corona Australis (154\,pc, Dzib et al. 2018), respectively.  Sz~102 and V853~Oph have GDR2 distances that differ significantly from the mean distance of their respective star-forming regions, Lupus and Ophiuchus, but also a high Astrometric Excess Noise. We follow \citet{Fang2018} in using the mean distance to Lupus~III and $\rho$~Oph for these two sources, 160\,pc and 138\,pc respectively.

We also collect literature spectral types (SpT), heliocentric radial velocities (v$_{\rm rad}$), stellar luminosities ($L_*$), mass accretion rates ($\dot{M}_{\rm acc}$), intrinsic X-ray luminosities (L$_X$), the total luminosity in the \oi{} 6300\,\AA{} line ($L_{\rm [OI]_{tot}}$) and its LVC contribution ($L_{\rm [OI]_{LVC}}$). We scale luminosities and accretion rates to the distances reported in Table~\ref{tab:targets}. Radial velocities are mostly taken from our high-resolution optical surveys \citep[][and Appendix~\ref{app:archive}]{Fang2018,Banzatti2019} and have a typical 1$\sigma$ uncertainty of 1\,km/s \citep{Pascucci2015}. The seven sources with an uncertainty greater than 3\,km/s can be divided in two groups\footnote{These  sources are DO~Tau, DR~Tau, HN~Tau, RU~Lup, VV~CrA, SR~21, and WaOph6, with a mean uncertainty in v$_{\rm rad}$ of 5\,km/s.}: high-accretors like DR~Tau with large veiling that reduces the depth of photospheric lines and extincted/optically-faint sources like SR~21 with low S/N spectra. SR~21 is also the only source in our sample with no detection of accretion-related optical emission lines \citep{Fang2018}, hence the mass accretion rate in Table~\ref{tab:targets} is an upper limit. Sz~102 has a nearly edge-on disk, therefore optical/infrared data largely underestimate its stellar luminosity  \citep{Alcala2017}. However, its stellar mass is known to be $\sim$1.6\,$M_\odot$ by modeling the  $^{12}$CO(3-2) Keplerian profile \citep{louvet2016}. As the source belongs to the Lupus~III star-forming region  which is $\sim 2$\,Myr old \citep{Alcala2017}, we report in Table~\ref{tab:targets} the stellar luminosity appropriate for such a star based on the evolutionary models of \citet{baraffe2015}. We use the stellar radius predicted from these models to calculate its mass accretion rate starting from the accretion luminosity reported in \citet{Fang2018}. Additional notes on complex systems are provided in Appendix~\ref{app:notes}.

\begin{longrotatetable}
\begin{deluxetable*}{ccccccccccccc}
\tablecaption{VISIR~2 and 1 sources with \oi{} 6300\,\AA{} detections. \label{tab:targets}}
\tablewidth{0pt}
\tablehead{
\colhead{ID}&\colhead{Target} & \colhead{2MASS} & \colhead{Dist} & \colhead{SpT}&  \colhead{v$_{\rm rad}$}& \colhead{Log $L_*$} & \colhead{Log $\dot{M}_{\rm acc}$} &   \colhead{Log $L_X$} &  \colhead{$n_{\rm 13-31}$} & \colhead{Log $L_{\rm [OI]_{tot}}$} & \colhead{Log $L_{\rm [OI]_{LVC}}$} & \colhead{Ref.} \\
\colhead{} & \colhead{} & \colhead{} & \colhead{(pc)} & \colhead{} & \colhead{(km/s)} &\colhead{($L_\odot$)} & \colhead{($M_\odot$/year)} & \colhead{($L_\odot$)} & \colhead{} & \colhead{($L_\odot$)} & \colhead{($L_\odot$)} & \colhead{}
}
\startdata
\multicolumn{13}{l}{VISIR 2:}\\
1 & GITau & J04333405+2421170 & 130.0 & M0.4 & 17.1 & -0.31 & -8.68 & -3.82 & -0.74 & -4.25 & -4.44 & 1,2,3,4 \\
2 & HNTau & J04333935+1751523 & 136.1 & K3 & 20.8 & -0.79 & -8.37 & -4.10 & -0.6 & -3.69 & -- & 1,2,3,4,5 \\
3 & AATau & J04345542+2428531 & 136.7 & M0.6 & 15.4 & -0.37 & -9.36 & -3.60 & -0.31 & -4.76 & -4.84 & 1,2,3,4 \\
4 & DOTau & J04382858+2610494 & 138.8 & M0.3 & 17.1 & -0.65 & -8.22 & -4.21 & -0.12 & -3.76 & -4.38 & 1,2,3,4 \\
5 & DRTau & J04470620+1658428 & 194.6 & K6 & 23.0 & -0.20 & -7.81 & -- & -0.33 & -4.16 & -4.27 & 1,2,3 \\
6 & V836Tau & J05030659+2523197 & 168.8 & M0.8 & 20.6 & -0.36 & -9.4 & -3.19 & -0.39 & -4.78 & -4.78 & 1,2,3,4 \\
7 & VWCha\tablenotemark{a} & J11080148-7742288 & 190.0 & K7 & 14.4 & 0.36 & -7.45 & -3.15 & -0.17 & -3.69 & -3.84 & 6,7,4 \\
8 & TWA3A & J11102788-3731520 & 36.6 & M4.1 & 12.3 & -1.19 & -10.15 & -4.44 &  -0.02 & -6.09 & -6.09 & 5,2,8\\
9 & GQLup & J15491210-3539051 & 151.2 & K6 & -2.9 & 0.17 & -7.38 & -3.35 & -0.22 & -4.04 & -4.04 & 9,2,5,4\\
10 & IMLup & J15560921-3756057 & 157.7 & K5 & -0.6 & 0.41 & -8.67 & -2.97 & -0.3 & -4.78 & -4.78 & 9,5,4\\
11 & RULup & J15564230-3749154 & 158.9 & K7 & 0.0 & 0.17 & -6.75 & -3.47 & -0.58 & -3.72 & -4.11 & 9,2,5,4\\
12 & RYLup & J15592838-4021513 & 158.4 & K2 & 0.8 & 0.27 & -8.58 & -2.62 & 0.68 & -4.63 & -4.63 & 9,2,5,10\\
13 & SR21 & J16271027-2419127 & 137.9 & F7 & -5.7 & 0.99 & $<$-8.39 & -3.50 & 1.82 & -4.8 & -4.8 & 5,4 \\
14 & RNO90 & J16340916-1548168 & 116.6 & G8 & -10.1 & 0.43 & -7.25 & -- & -0.5 & -4.16 & -4.16 & 2,5 \\
15 & WaOph6 & J16484562-1416359 & 123.4 & K7 & -7.6 & -0.12 & -7.34 & -- & -0.46 & -4.84 & -- & 2,5 \\
16 & V4046Sgr\tablenotemark{a} & J18141047-3247344 & 72.3 & K5+K7 & -6.2 & -0.23 & -9.22 & -3.51 & 0.87 & -5.45 & -5.45 & 11,12,13,14,7 \\
17 & SCrA A+B\tablenotemark{a} & J19010860-3657200 & 152.3 & K6 & 2.5 & 0.25 & -7.42 & -3.36 & 0.19 & -3.51 & -- & 2,5,15 \\
18 & TYCrA\tablenotemark{a} & J19014081-3652337 & 136.5 & B9 & -4.6 & 1.41& -8.04& -2.84\tablenotemark{b} & --\tablenotemark{b} & -4.02 & -- & 16,17,18,15,7 \\
19 & TCrA\tablenotemark{a} & J19015878-3657498 & 154.0 & F0 & 1.1 & 1.46 & -7.94 & -5.44 & 0.91 & -3.72 & -- & 7,19,18,14 \\
20 & VVCrA S\tablenotemark{a} & J19030674-3712494 & 148.8 & K7 & -5.7 & 0.33 & -6.42 & -- & 0.0 & -3.05 & -3.77  & 2,5 \\
\hline
\multicolumn{13}{l}{VISIR 1:}\\
21 & LkCa15 & J04391779+2221034 & 158.2 & K5.5 & 18.7 & -0.11 & -8.74 & -3.00 & 0.64 & -5.05 & -5.05 & 5,1,20 \\
22 & TWHya & J11015191-3442170 & 60.0 & M0.5 & 13.6 & -0.63 & -8.67 & -3.26 & 0.68 & -4.95 & -4.95 & 5,4,3\\
23 & CSCha\tablenotemark{a} & J11022491-7733357 & 175.4 & K2 & 15.5 & 0.24 & -8.21 & -3.02 &  2.84 & -4.9 & -4.9 & 6,7,4 \\
24 & VZCha & J11092379-7623207 & 191.2 & M0.5 & 19.0 & -0.1 & -7.18 & -3.80 & -1.07 & -4.69 & -4.69 & 6,21,4,22 \\
25 & TCha & J11571348-7921313 & 109.3 & G8 & 15.8 & 0.48 & -8.11 & -3.10 & 1.33 & -4.64 & -4.64 & 23,7,24,4 \\
26 & MPMus & J13220753-6938121 & 98.6 & K1 & 10.7 & -1.13 & -8.56 & -3.46 & 0.05 & -4.82 & -4.82 & 22,7,14\\
27 & Sz73 & J15475693-3514346 & 156.1 & K7 & -3.6 & -0.34 & -8.53 & -- & -0.1 & -4.23 & -4.81 & 9,5 \\
28 & Sz102 & J16082972-3903110 & 160.0 & K2 & 12.0 & -0.5 & -9.15 & -4.52 & 0.57 & -3.75 & -- & 5,7,4 \\
29 & V853Oph & J16284527-2428190 & 138.0 & M2.5 & -5.8 & -0.3 & -8.08 & -3.01 & -0.45 & -4.56 & -4.76 & 5,4 \\
30 & DoAr44 & J16313346-2427372 & 145.3 & K2 & -4.5 & -0.02 & -8.04 & -3.34 & -0.45 & -4.81 & -4.81 & 5,4 \\
31 & RXJ1842.9-35 & J18425797-3532427 & 153.2 & K3 & -0.9 & -0.22 & -8.51 & -3.14 & 0.64 & -4.34 & -4.41 & 5,25 \\
\enddata
\tablecomments{Stellar luminosities, mass accretion rates, X-ray luminosities, and \oi{} 6300\,\AA{} luminosities are scaled to the distances given in this table. Except for RXJ1842.9-35, all $L_X$ are intrinsic, i.e. corrected for absorption, and representative for the energy band $0.3-10$\,keV. For  RXJ1842.9-35 the only $L_X$ available is from $ROSAT$ PSPC ($0.1-2.4$\,keV).
The spectral index for TWA3A is actually that for the  A+B system and it is derived from the {\it Spitzer} IRAC 8\,\micron{} and MIPS 24\,\micron{} photometry. Uncertainties in v$_{\rm rad}$ are discussed in Section~\ref{sec:sample}.}
\tablenotetext{a}{Additional notes on complex systems are provided in Appendix~\ref{app:notes}.}
\tablenotetext{b}{The X-ray luminosity of TY~CrA is likely dominated by the three later type companions, no infrared index can be computed due to uneven nebular background emission (see additional info in Appendix~\ref{app:notes}).}
\tablerefs{1. \citet{HerczegHillenbrand2014}; 2. \citet{Banzatti2019}; 3. \citet{Simon2016}; 4 \citet{Guedel2010}; 5. \citet{Fang2018}; 6. \citet{Manara2017}; 7. This work; 8. \citet{Kastner2016}; 9. \citet{Alcala2017}; 10. \citet{Dionatos2019}; 11. \citet{Rodriguez2010} ; 12. \citet{Rosenfeld2013}; 13. \citet{Curran2011}; 14. \citet{Sacco2012}; 15. \citet{ForbrichPreibisch2007}; 16. \citet{Casey1993}; 17. \citet{Vioque2018}; 18. \citet{Dong2018}; 19. \citet{Cazzoletti2019}; 20. \citet{SkinnerGuedel2017}; 21. \citet{Torres2006}; 22. \citet{Rigliaco2013}; 23 \citet{Schisano2009}; 24. \citet{Cahill2019}; 25. \citet{Pascucci2007} }
\end{deluxetable*}
\end{longrotatetable}

The majority of our sources have \oi{} 6300\,\AA{} LVC emission, only for 6 disks the entire emission can be attributed to jets (HN~Tau, Wa~Oph6, SCrA~A+B, TY~CrA, T~CRA, and Sz~102, see Table~\ref{tab:targets}). Note that in several of these cases it is the close to edge-on view that precludes us from kinematically separating the LVC from the HVC, see also Appendix~\ref{app:archive}. Among the LVC, 9 have a BC {\it and} a NC, while the remaining 16 have either a BC {\it or} a NC. Because of the lower spectral resolution and typically lower S/N of the VISIR spectra, we cannot distinguish additional components within the \neii{} LVC. Hence, moving forward, we will not discuss the  \oi{} NC and BC separately but rather combine their luminosities  as in Table~\ref{tab:targets} and only separate the \oi{} LVC from the HVC.

To characterize the level of dust depletion in the inner disk, we also calculate the infrared spectral index $n_{\rm 13-31}$ defined as in \citet{furlan2009}, and provide it in Table~\ref{tab:targets}. To this end we retrieved fully reduced medium-resolution ($R\sim 700$) {\it Spitzer} archival spectra from \citet{Pontoppidan2010} or from the online CASSIS database \citep{Leb2015} and used the same wavelength ranges as in \citet{Banzatti2019} to compute the mean flux densities at the relevant wavelengths. The CASSIS database only offers low-resolution {\it Spitzer} spectra of MP~Mus and V836~Tau, so we use the wavelength ranges adopted in \citet{furlan2009}, appropriate for the lower spectral resolution. Finally, the high-resolution {\it Spitzer}  spectrum from TWA3A is of poor quality, with a large discontinuity between the short and long wavelength modules. Hence, for this source we use the {\it Spitzer} IRAC and MIPS photometry closest to the 13 and 31\,\micron{} wavelengths to calculate the spectral index  in Table~\ref{tab:targets}. Spectral indices greater than $\sim 0$ point to dust depletion and the presence of inner cavities \citep{furlan2009}. Indeed, the well-known disks with dust cavities around SR~21, LkCa~15, TW~Hya, CS~Cha, and T~Cha \citep[e.g.,][]{vanderMa2016} have relatively large and positive spectral indices.

In addition to this sample of 31 disks observed at high spectral resolution at optical {\it and} infrared wavelengths, we will include in Section~\ref{sect:correlations} five more disks
with $n_{\rm 13-31} \gtrsim 1$, \oi{} detections but no HVC, and medium-resolution (R$\sim 700$) infrared spectra from {\it Spitzer}/IRS, see Table~\ref{tab:spitzer_neii}. As demonstrated in Appendix~\ref{app:comparison} using our Table~\ref{tab:targets} sources, disks with $n_{\rm 13-31} \gtrsim 1$ and no \oi{} HVC have {\it Spitzer} \neii{} fluxes well within a factor of two of the VISIR ones. Hence, such disks can be used to expand the high-resolution mid-infrared sample.

\begin{figure*}[ht!]
\plotone{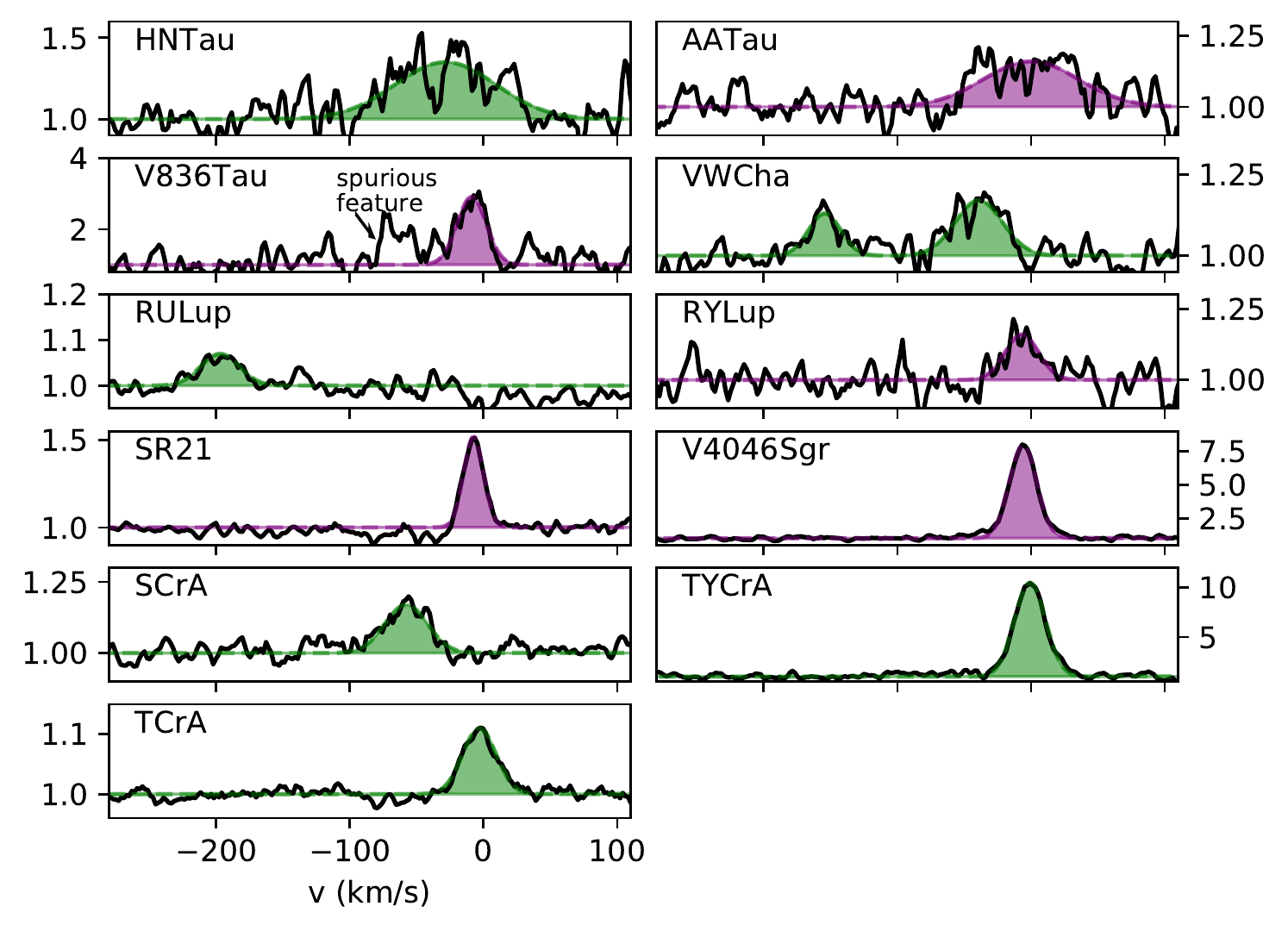}
\caption{VISIR~2 spectra with a \neii{} 12.8\,\micron{} detection. For visualization purposes, we applied a boxcar smoothing of 3 velocity elements. For SCrA we show the spectrum from the fainter B component, see Figure~\ref{fig:scra} for the other component. The best fit Gaussian profiles are colored in green for the HVC and purple for the LVC following the assigments in Table~\ref{tab:neii}. \label{fig:visir2}}
\end{figure*}

\section{\neii{} fluxes and upper limits} \label{sec:NeII}
We start from the fully reduced VISIR 2.0 spectra which were corrected for telluric absorption,
referenced to the heliocentric frame, and normalized to the continuum. We shift the spectra to the stellocentric reference frame using the stellar radial velocities reported in Table~\ref{tab:targets}. To determine if the \neii{} line at 12.81\,\micron{} is detected, we calculate the root-mean-square (rms) per pixel in two spectral regions that are free of emission, short-ward and long-ward of the transition itself. We consider the line to be detected if multiple wavelengths close to 12.81\,\micron{} have emission above three times the rms.

\begin{figure}
\plotone{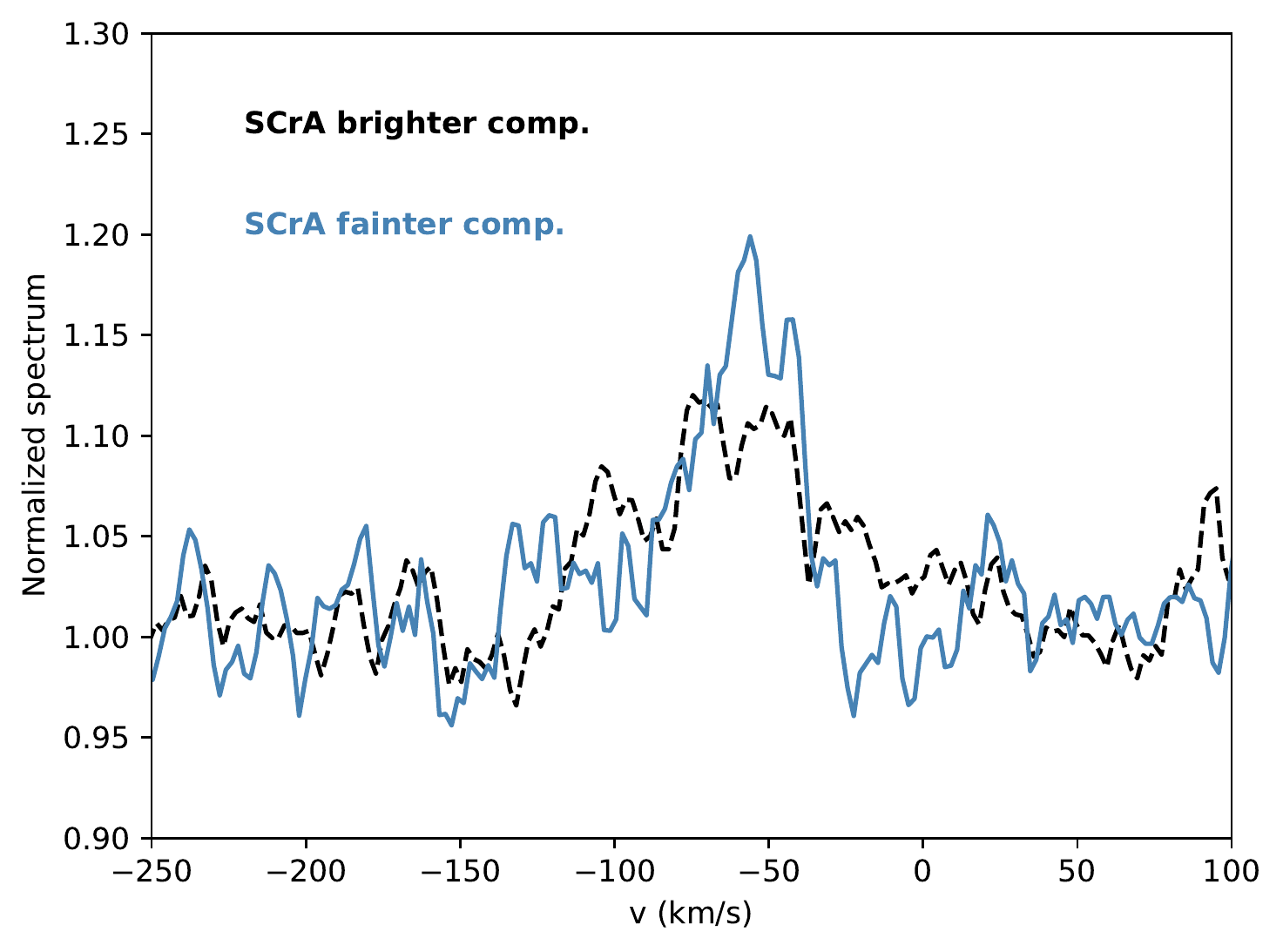}
\caption{VISIR~2 spectra of the SCrA binary system. Note that for both companions the detected \neii{} emission is significantly blueshifted from the stellar velocity, here at zero. \label{fig:scra}}
\end{figure}

Eleven out of the 20 VISIR~2 sources listed in Table~\ref{tab:targets} have a \neii{} detection with HN~Tau, V836~Tau, VW~Cha, RY~Lup, SCrA, and TY~CrA being new discoveries\footnote{Wa~Oph6 has only a tentative 2$\sigma$ detection at $\sim -100$\,km/s, hence it is not included among the detections.} --- see Figure~\ref{fig:visir2}. Note that the bump at $\sim -65$\,km/s in the spectrum of V836~Tau is a spurious feature caused by the removal of high-frequency fringing and of a CO$_2$ telluric line while the emission close to the stellar velocity is real. SCrA is a similar spectral type binary separated by 1.3\arcsec{}  \citep{Sullivan2019} and we detect blueshifted \neii{} emission of similar intensity from the optically brighter A and the fainter B star --- see Figure~\ref{fig:scra}. Because stellar properties obtained from the optical spectra are for the A+B component \citep{Fang2018}, in the following sections we will use the sum of the \neii{} emission for this source.

If \neii{} emission is detected, we follow an approach similar to that adopted for the oxygen forbidden lines in that we fit the minimum number of Gaussian profiles to reproduce the observed line (e.g., \citealt{Simon2016} and Appendix~\ref{app:archive})\footnote{We have also tested a different approach whereby we  fit a Gaussian profile and calculate the EW in the wavelength range given by the Gaussian centroid $\pm$ 3$\sigma$ where $\sigma$ is the standard deviation of the Gaussian. Due to the poor S/N of the spectra, we found that this method tends to underestimate the EW, hence we prefer the Gaussian fitting.}.
Except for VW~Cha, one Gaussian is sufficient to reproduce the observed profiles --- see Figure~\ref{fig:visir2}. Gaussian FWHMs, centroids (v$_{\rm c}$), and EWs for the detected lines are given in Table~\ref{tab:neii}. Uncertainties in v$_{\rm c}$ and FWHM range from a maximum of 5\,km/s for sources with a low S/N detection, such as HN~Tau and AA~Tau, to below 1\,km/s for those with a strong detection, such as SR~21 and TY~CrA. However,  as the 1$\sigma$ uncertainty on the stellar radial velocity is typically 1\,km/s (Section~\ref{sec:sample}),  even for sources with high S/N spectra the \neii{} centroid is not known to better than 1\,km/s.

As with the forbidden oxygen lines  \citep{Simon2016}, we call a component HVC (LVC) if the centroid is more (less) blueshifted than 30\,km/s. Following this classification we would have 8 LVCs and 6 HVCs in our VISIR~2 sample. However, because the HVC (jet emission) centroid is anti-correlated with disk inclination (e.g., \citealt{Banzatti2019} for the \oi{} 6300\,\AA{} line), the HVC from highly inclined disks could show blueshifts smaller than 30\,km/s and be classified as LVC, if inclination is not taken into account. We have three such inclined systems in our VISIR~2 sample (Table~\ref{tab:neii}). For HN~Tau the re-assigment of the \neii{} LVC into HVC is in line with \citet{Fang2018} who find that even the modestly blueshifted ($\sim$\,-10\,km/s) optical forbidden line components for HN~Tau have line ratios that are more compatible with HVCs than with LVCs. In the case of TY~CrA, the low critical density \sii{} 6731\,\AA{} line, which is a  well established jet diagnostic  \citep[e.g.,][]{Hartigan1995,Natta2014}, peaks at the stellar velocity (Figure~\ref{fig:tycra}). Finally, for T~CrA  Whelan et al. in prep. analyze the 2D spectra from several optical forbidden lines and discover a jet aligned with the plane of the sky and thus with radial velocities close to zero.
We note that using de-projected velocity centroids and a minimum shock velocity of $\sim 30$\,km/s as inferred for several jets \citep[e.g.,][]{Hartigan1994}, would result in the same HVC/LVC classification, except for TY~CrA. For this complex system (Appendix~\ref{app:notes}), assuming at face value the inclination of the eclipsing binary as that of the disk, would give a de-projected centroid velocity of only -9\,km/s. However, as mentioned above, the detection of the \sii{} 6731\,\AA{} line with no blueshift with respect to the stellar velocity points to a jet in the plane of the sky.

In the absence of \neii{} emission, we follow \citet{Sacco2012} and provide an upper limit equal to $5 \times  \rm{rms} \times \sqrt{FWHM \, \delta {\it v}}$  with $\delta v$ being the line width of a velocity bin ($\sim 2$\,km/s) and the FWHM taken to be 20\,km/s for comparison with photoevaporative winds \citep[e.g.,][]{ErcolanoOwen2010}. These EW upper limits are given as negative values in  Table~\ref{tab:neii}.

To convert EWs into fluxes we use the flux density near the \neii{} line measured on low- or medium-resolution {\it Spitzer} spectra  \citep[e.g.,][]{Pontoppidan2010,Rigliaco2015}. Exceptions are TWA3A, for which we use WISE/W3 broad-band photometry,  and T~CrA, whose {\it Spitzer} flux density at 12.81\,\micron{} is twice as large than that measured by VISIR~1 \citep{Sacco2012} probably due to extended emission within the large slit of {\it Spitzer}/IRS. While the absolute flux calibration of the {\it Spitzer} spectra is accurate to $\sim 10$\% \citep[e.g.][]{Pascucci2007}, annual/decadal mid-infrared variability larger than $\sim$\,20\% is common in young stars \citep[e.g.,][]{Espaillat2011,Kospal2012}. Therefore, mid-infrared variability, which is mostly unknown for our sources, is likely to be the dominant uncertainty in the \neii{} luminosities reported in Table~\ref{tab:neii}.

The \neii{} data for the VISIR~1 sample are collected from the literature and also provided in Table~\ref{tab:neii} for completeness. We have assigned a classification (Type) to the detections following the work on the forbidden optical lines \citep{Simon2016}.

\begin{deluxetable*}{cccccccccc}
\tablecaption{VISIR \neii{} detections and upper limits. \label{tab:neii}}
\tablewidth{0pt}
\tablehead{
\colhead{ID} & \colhead{Target} & \colhead{FWHM} & \colhead{v$_{\rm c}$} & \colhead{EW}  & \colhead{$i$}&\colhead{F$_{\rm cont}$}   & \colhead{Log$L_{\rm [NeII]}$} & \colhead{Type} & \colhead{$i$,F$_{\rm cont}$,[NeII]}\\
\colhead{} & \colhead{} & \colhead{(km/s)} & \colhead{(km/s)} &  \colhead{(\AA)} & \colhead{($^\circ$)}& \colhead{(Jy)}  & \colhead{($L_\odot$)}  & \colhead{} & \colhead{Ref.}
}
\startdata
\multicolumn{10}{l}{VISIR 2:}\\
1 & GITau &  &  &-1.59 & & 0.76  & $<$ -5.93 & &1\\
2 & HNTau\tablenotemark{b} & 89.03 & -28.42 & 14.05 & 75& 0.95 & -4.85 & LVC$\rightarrow$HVC\tablenotemark{c} & 2,3\\
3 & AATau & 83.96 & 0.01 & 6.01 & 59&0.31 & -5.7 & LVC & 4,3\\
4 & DOTau &  &  & -2.24 & & 1.89 & $<$ -5.33 & & 3 \\
5 & DRTau &  &  & -2.0 & & 1.88 &  $<$ -5.09 & & 3 \\
6 & V836Tau & 25.67 & -8.89 & 22.31 & 61& 0.1 & -5.44 & LVC & 5,3\\
7 & VWCha & 41.83 & -39.04 & 3.3 & 45& 0.73 & -5.3 & HVC & 6,3\\
7  & & 28.3 & -154.38 & 1.7 & 45 &0.73 & -5.59 & HVC & 3\\
8 & TWA3A &  &  & -2.75 & & 0.89 & $<$-6.73 & & 7\\
9 & GQLup &  &  & -1.72 & & 0.49 & $<$-5.96 & & 3\\
10 & IMLup &  &  & -1.08 & & 0.48 &  $<$-6.13 & & 3\\
11 & RULup  & 33.9 & -197.04 & 1.0  & 35& 4.29 & -5.21 & HVC & 8,3\\
12 & RYLup & 29.68 & -6.91 & 2.12 & 67& 0.76 & -5.64 & LVC & 9,1\\
13 & SR21\tablenotemark{d} & 17.44 & -7.41 & 4.14 & 18& 2.1 & -5.02 & LVC & 5,1\\
14 & RNO90 &  &  & -0.27 & &2.05 &  $<$-6.36 & & 3 \\
15 & WaOph6 &  &  & -0.93 & &0.82 & $<$-6.18 & & 3\\
16 & V4046Sgr\tablenotemark{d} & 24.0 & -5.97 & 76.07 & 35&0.37 & -5.08 & LVC & 10,1\\
17 & SCrA~A & 81.97 & -60.51 & 3.71 & 10&4.61 & -4.64 & HVC & 8,1 \\
17 & SCrA~B& 37.34 & -57.83 & 2.95 & 10&4.61 & -4.74 & HVC & 1 \\
18 & TYCrA & 27.26 & -0.82 & 117.65 &  85&1.5 & -3.73 & LVC$\rightarrow$HVC\tablenotemark{c} & 11,12\\
19 & TCrA & 29.61 & -3.11 & 1.44 & 90&2.8 & -5.26 & LVC$\rightarrow$HVC\tablenotemark{c} & 13,14 \\
20 & VVCrA &  &  & -0.02 & &27.1 &  $<$-6.16 & & 1 \\
\hline
\multicolumn{10}{l}{VISIR 1:}\\
21 & LkCa15 &  &  & & & & $<$-5.41  & & 14\\
22 & TWHya & 16.3 & -4.8 & & 7& & -5.35 & LVC & 15,16 \\
23 & CSCha &  27.0 & -3.3 & & 11&  &  -4.66 & LVC & 17,18 \\
24 & VZCha &  &  & & & &  $<$-5.58  &  & 19\\
25 & TCha &  42.0 & -4.0 & & 75&  & -5.09 & LVC & 20,18 \\
26 & MPMus & 15.9 & -4.4 & & 30&  & -5.48 & LVC & 21,14 \\
27 & Sz73 & 60 & -99.0  & & 48& &  -5.08 &HVC & 22,18 \\
28 & Sz102 & &  & & & &   $<$-5.40 & & 18 \\
29 & V853Oph &  26.5 & -35.8  & & 54& & -5.66& HVC & 5,19 \\
30 & DoAr44 & &  & & & &   $<$-5.7 & & 14 \\
31 & RXJ1842.9-35 & &  & & &  &  $<$-5.83 & & 14\\
\enddata
\tablecomments{Negative EW values are used to indicate upper limits. Disk inclinations are provided for sources with a \neii{} detection to evaluate de-projected centroid velocities. References are given for the disk inclination, the continuum flux density (F$_{\rm cont}$) for VISIR~2 data, and the \neii{} properties for VISIR~1 data.  }
\tablenotetext{b}{Dataset acquired on 26-11-2017, the emission is also detected in the 28-12-2017 dataset at a slightly lower S/N}
\tablenotetext{c}{Highly inclined disks, see Section~\ref{sec:NeII} and Appendix~\ref{app:notes} for details on the re-assignments.}
\tablenotetext{d}{SR~21 and V4046~Sgr were observed in two and four slit orientations, respectively. The \neii{} emission is detected in all exposures with similar shape and intensity. Here, we provide results from the first exposures, slits oriented N-S.}
\tablerefs{1. this work (using {\it Spitzer}/IRS spectra from either the CASSIS database or from \citealt{Pontoppidan2010}); 2. \citet{Simon2017}; 3. \citet{Rigliaco2015}; 4. \citet{Loomis2017};
5. \citet{Tripathi2017}; 6. \citet{McClure2015};
7. \citet{Kellogg2017}; 8. \citet{Pontoppidan2011};
9. \citet{Francis2020}; 10. \citet{Rosenfeld2013};
11. \citet{Vanko2013}; 12. \citet{Boersma2009};
13. Whelan et al. in prep.; 14. \citet{Sacco2012};
15. \citet{Andrews2016}; 16. \citet{Pascucci2011};
17. \citet{Hendler2020}; 18. \citet{PascucciSterzik2009};
19. \citet{Baldovin2012}; 20. \citet{Schisano2009};
21. \citet{Kastner2010}; 22. \citet{Ansdell2016}}
\end{deluxetable*}

\begin{figure*}[ht!]
\plotone{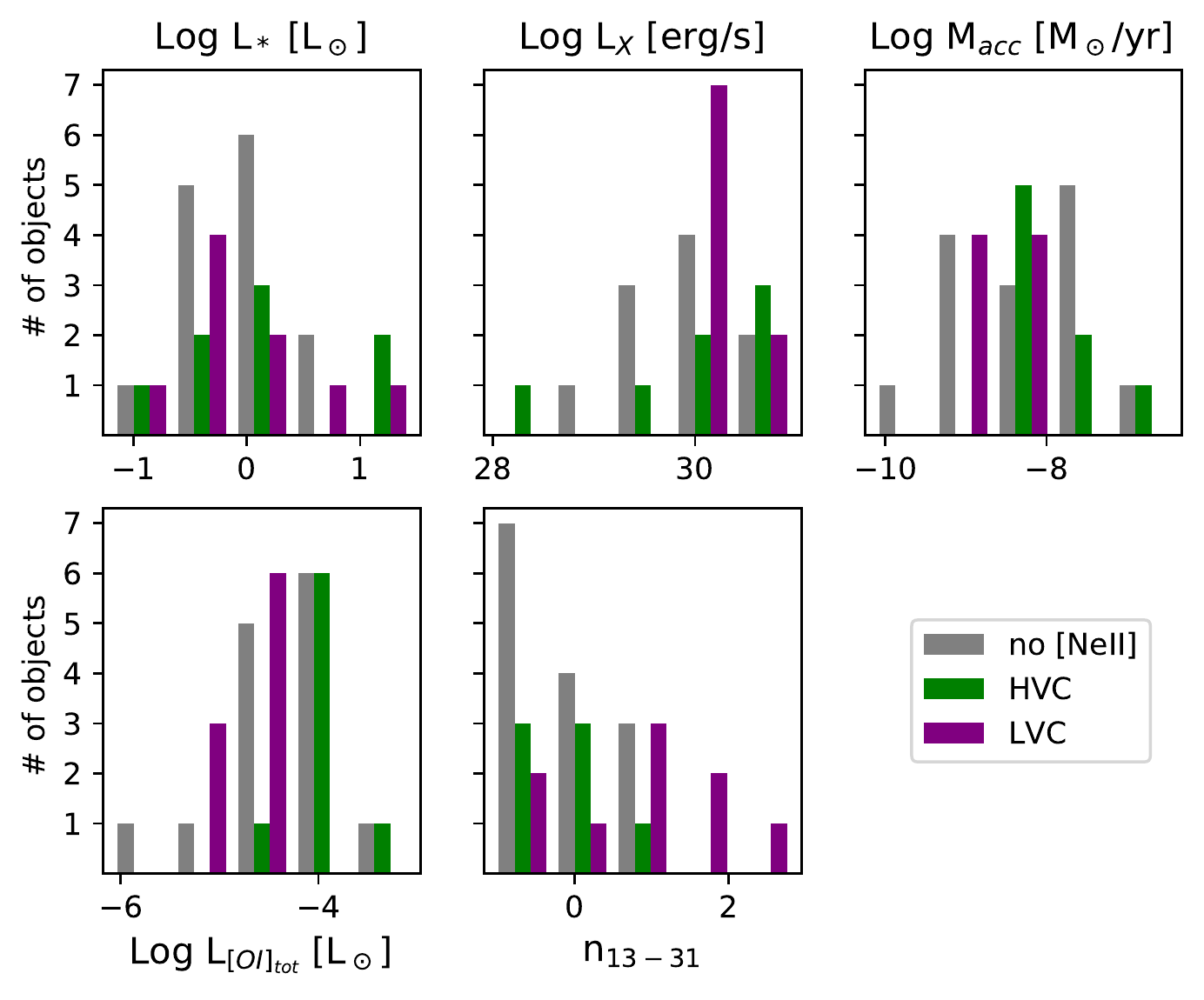}
\caption{Histograms showing the distribution of \neii{} HVC (green) and LVC (purple) detections in comparison to that of non-detections (grey) for the sources in Table~\ref{tab:targets}. Note that the total number of objects varies from panel to panel. When \neii{} emission is detected, HVCs dominate in sources with large $\dot{M}_{\rm acc}$
and large $L_{[OI]_{\rm tot}}$
while LVCs in sources with low $\dot{M}_{\rm acc}$, low $L_{[OI]_{\rm tot}}$, and large $n_{\rm 13-31}$.
\label{fig:histoNeII}}
\end{figure*}

\section{Results}\label{sec: results}
The combined VISIR sample has a total of 17 \neii{} 12.81\,\micron{} detections out of the 31 targets listed in Table~\ref{tab:targets}.
Nine sources present a \neii{} LVC while eight show a HVC.
All disks with a \neii{} HVC(LVC) detection also have a HVC(LVC) detected in the \oi{} 6300\,\AA{} line. Figure~\ref{fig:histoNeII} shows the distribution of \neii{} HVC and LVC detections, as well as non-detections, for the stellar/disk properties summarized in Table~\ref{tab:targets}. When \neii{} emission is detected, sources with $\dot{M}_{\rm acc} > 10^{-8}$\,M$_\odot$/yr have only a HVC. The same is true for systems displaying a large total \oi{} 6300\,\AA{} luminosity ($L_{\rm [OI]_{tot}} > 5.4 \times 10^{-5}$\,L$_\odot$).  On the contrary, \neii{} LVC is preferentially detected toward sources with low  $\dot{M}_{\rm acc}$, low \oi{} 6300\,\AA{} luminosity, and high infrared spectral index ($n_{\rm 13-31}> 0.5$).

In the following, we will first explore if the known correlations between the \oi{} luminosities and stellar/disk properties apply to the \neii{} line luminosities (Section~\ref{sect:correlations}). Next, we will compare the \oi{} and \neii{} line profiles for individual kinematic components to identify possible trends between line centroids and FWHM (Section~\ref{sect:profiles}).

\begin{figure*}[ht!]
\plotone{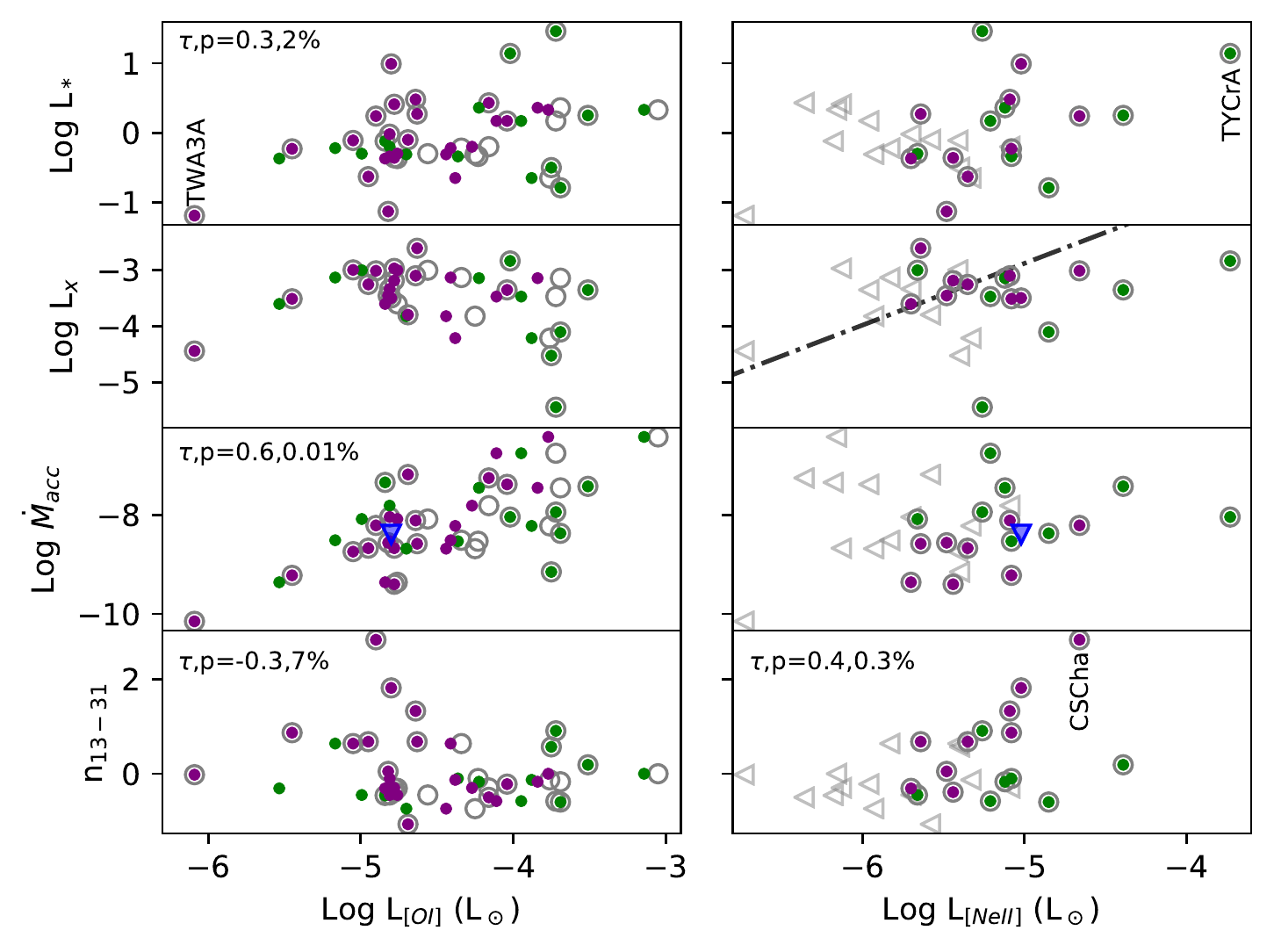}
\caption{Left panels:  Stellar luminosity in L$_\odot$ (upper), X-ray luminosity in L$_\odot$ (second),
mass accretion rate in M$_\odot$/yr (third), and infrared spectral index (bottom) vs the \oi{} 6300\,\AA{} luminosity --- empty grey symbols for the total luminosity while filled purple(green) symbols for the LVC(HVC). SR~21, the only source with an upper limit on the mass accretion rate, is plotted with a blue downward triangle. According to the  Kendall's $\tau$ test the LVC \oi{} luminosity is likely correlated with the stellar luminosity (positive), mass accretion rate (positive), and spectral index (negative).  Right panels: Same but for the \neii{} luminosity. The black dot-dashed line in the third panel shows the expected relation between X-ray and \neii{} luminosity \citep{HollenbachGorti2009}.
There is a low probability that the infrared spectral index and the \neii{} LVC luminosity are not positively correlated.
Complete statistics for the LVC and HVC are given in Table~\ref{tab:stat}. \label{fig:oineii}}
\end{figure*}

\subsection{Correlations between line luminosities and stellar/disk properties}\label{sect:correlations}
Recent medium- ($\Delta v \sim 35$\,km/s) and high-resolution optical ($\Delta v \sim 7$\,km/s) surveys have  established that there exists a number of correlations between individual \oi{} 6300\,\AA{} component's properties with  stellar and disk properties. In particular, \cite{Nisini2018} showed that the LVC and HVC luminosities correlate better with accretion luminosity and mass accretion rate rather than with stellar luminosity and mass.
\cite{Banzatti2019} further separated LVC into single and double components and discovered that the line EW of single components (without any jet emission) is anti-correlated with the infrared spectral index, i.e. disks with inner dust depletion have lower \oi{} 6300\,\AA{} EWs. In relation to the X-ray luminosity, previous surveys from our group did not identify any correlation between $L_X$ and the  $L_{\rm [OI]_{LVC}}$ \citep{Rigliaco2013,Simon2016}. In contrast, \citet{McGinnis2018} reported a positive correlation but found that it was driven by the stronger correlation between the $L_{\rm [OI]_{LVC}}$ and  $L_*$ in their NGC~2264 sample. The same study reported
no correlation  between
$L_X$ and $L_{\rm [OI]_{HVC}}$.

\begin{deluxetable}{lcccc}
\tablecaption{Summary of the {\tt cenken} Kendall's $\tau$ tests.\label{tab:stat}}
\tablewidth{0pt}
\tablehead{
\colhead{Quantity} & \multicolumn2c{LVC} & \multicolumn2c{HVC} \\
\cline{2-3}
\cline{4-5}
\colhead{} & \colhead{$L_{\rm [OI]}$} & \colhead{$L_{\rm [NeII]}$} & \colhead{$L_{\rm [OI]}$} & \colhead{$L_{\rm [NeII]}$}
}
\startdata
$L_*$ &  0.3(2) & \dem{0.1(72)}  & \dem{0.2(34)} & \dem{0.1(48)} \\
$L_X$ & \dem{0(81)} & \dem{0.1(38)} & \dem{-0.2(45)} &  \dem{0.1(73)}\\
$\dot{M}_{acc}$ & 0.6(0.01) & \dem{0(93)}  & 0.4(6) & \dem{0.1(68)}  \\
$n_{\rm 13-31}$ & -0.3(7) & 0.4(0.3) & \dem{0.2(32)} &  \dem{0(87)}\\
\enddata
\tablecomments{The first entry ($\tau$) runs from -1 to 1 and indicates the direction of the correlation while the value in parenthesis (p) is the percent probability that the two quantities are uncorrelated. Entries with probabilities larger than 10\% are greyed out. }
\end{deluxetable}

The left panels of Figure~\ref{fig:oineii} show how stellar luminosities ($L_*$), intrinsic X-ray luminosities ($L_X$), mass accretion rates ($\dot{M}_{\rm acc}$), and infrared spectral indices ($n_{\rm 13-31}$) relate to the  \oi{} 6300\,\AA{} luminosity ($L_{\rm [OI]}$) for our combined VISIR sample. We use the {\tt cenken} $R$ routine to compute the nonparameteric Kendall's $\tau$ correlation coefficient and associated probability between the aforementioned stellar/disk properties
and the \oi{} LVC and HVC luminosities (see Table~\ref{tab:stat}).   {\tt cenken} properly accounts for individual upper limits (censored data) which is particularly useful when there are many non-detections as for the \neii{} line, see below. For the LVC, we find that our restricted sample recovers the same trends reported in the literature: likely positive correlations between $L_*$ and  $\dot{M}_{\rm acc}$ with $L_{\rm [OI]_{LVC}}$ (with $\dot{M}_{\rm acc}$ displaying a higher degree of correlation) and a likely negative correlation between $n_{\rm 13-31}$ and  $L_{\rm [OI]_{LVC}}$. For the HVC, our restricted sample  only recovers the known correlation between its component \oi{} luminosity and $\dot{M}_{\rm acc}$.

\begin{figure}[h]
\plotone{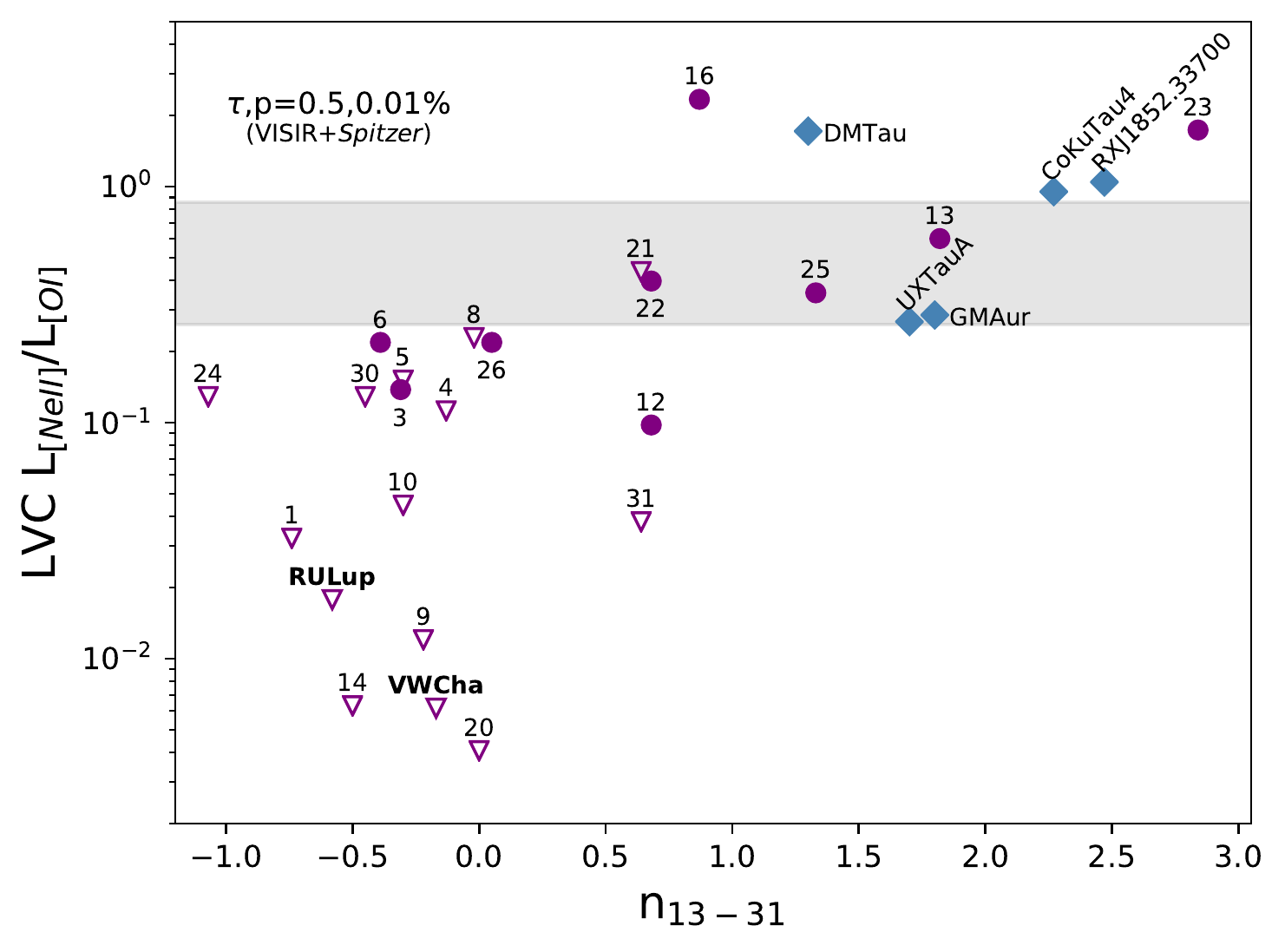}
\caption{LVC line ratios  vs infrared spectral index. Purple circles and down-pointing triangles (non-detections) are for the VISIR sample (source ID as in Table~\ref{tab:targets}). RU~Lup and VW~Cha (in bold) have a HVC but no LVC detection. Blue diamonds are for the {\it Spitzer} sources identified in Appendix~\ref{app:comparison}. The grey band shows the range of predicted line ratios from \citet{ErcolanoOwen2010}, with the upper bound multiplied by a factor of 2 to account for the possible underestimation of the $L_{\rm [NeII]}$. \label{fig:neii_oi}}
\end{figure}

In the right panels of Figure~\ref{fig:oineii} we show how the same stellar/disk properties relate to the \neii{} 12.81\,\micron{} luminosities ($L_{\rm [NeII]}$) and test for correlations  on the LVC and HVC detections and upper limits (14 over 31) using {\tt cenken}  (Table~\ref{tab:stat}). The test only identifies a significant trend between the infrared spectral index and the \neii{} LVC luminosity: a low probability that $n_{\rm 13-31}$ and $L_{\rm [NeII]}$ would be so highly correlated through chance alone suggests that disks with inner dust depletion do have higher \neii{} 12.81\,\micron{} LVC luminosities\footnote{Even when excluding CS~Cha, which has the highest $n_{\rm 13-31}$ and $L_{\rm [NeII]_{LVC}}$, {\tt cenken} returns a Kendall's $\tau$ probability of only 1\%} that the two quantities are uncorrelated. This trend is not driven by S/N as exposure times were similar among sources with low and high $n_{\rm 13-31}$. Note that this is opposite to the trend between  $n_{\rm 13-31}$ and the \oi{} LVC luminosity.

\begin{figure*}[ht!]
\plottwo{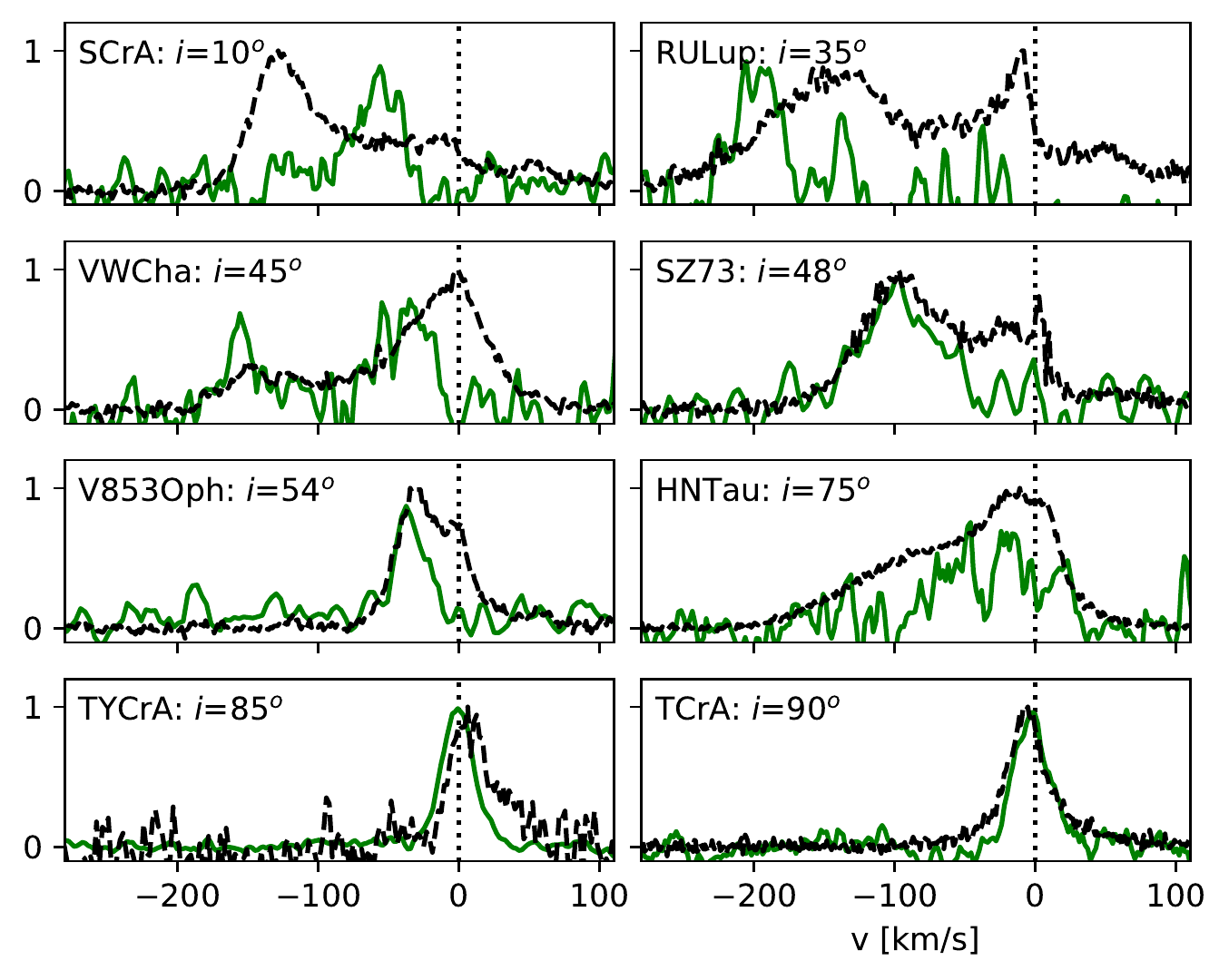}{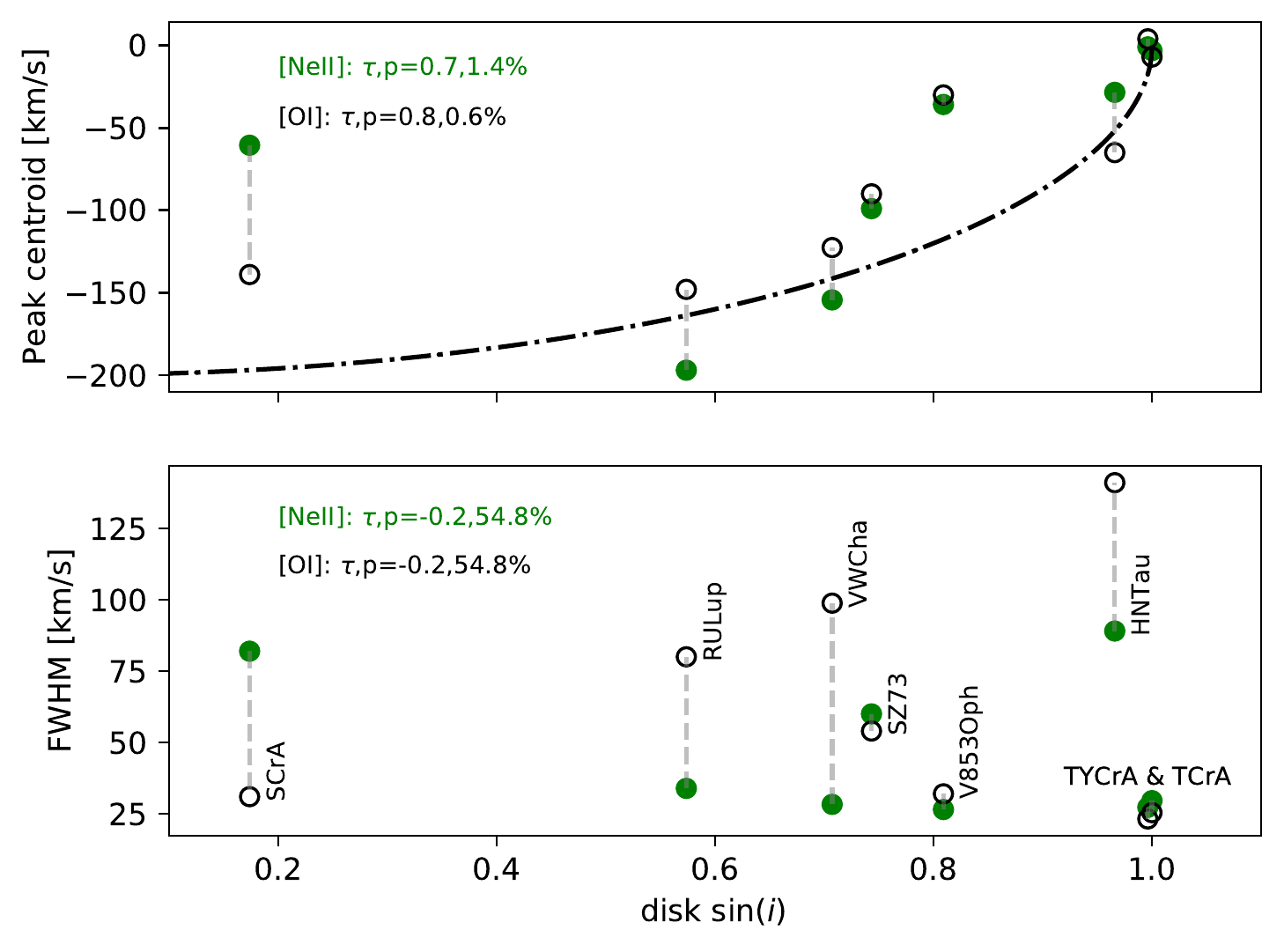}
\caption{Left panels: Comparison of normalized \neii{} 12.81\,\micron{} (green) and \oi{} 6300\,\AA{} (black) profiles for disks with a HVC component. Sources are ordered by increasing disk inclination. Right panels: centroid and FWHM vs disk inclination. The dot-dashed line in the upper panel shows the projected radial velocity for a micro-jet with intrinsic velocity of -200\,km/s and perpendicular to the disk midplane. \label{fig:hvcprof}}
\end{figure*}

To better highlight this result, we show in Figure~\ref{fig:neii_oi} the ratio of the forbidden line LVC luminosities ($L_{\rm [NeII]}$/$L_{\rm [OI]}$) vs the infrared spectral index ($n_{\rm 13-31}$). For the entire VISIR sample, \neii{} detections (purple circles) and non-detections (purple downward triangles), the  Kendall's $\tau$ probability that the two quantities are uncorrelated is only 0.6\%. Adding the five {\it Spitzer}/IRS sources with no jet emission and whose fluxes would likely be recovered with the narrow slits of VISIR (Appendix~\ref{app:comparison}, blue diamonds) further decreases the probability to 0.01\%.
Except V4046~Sgr (observed with VISIR) and DM~Tau (observed with {\it Spitzer}), the data suggest that the $L_{\rm [NeII]}$/$L_{\rm [OI]}$ ratio increases for $n_{\rm 13-31} \ge 1.5$, i.e. for more depleted dust cavities, while it is  at most $\sim 0.4$ for lower spectral indices. However, note that the majority of the sources with no inner dust depletion, while having strong LVC \oi{} emission, are not detected in the \neii{}. Even when their infrared spectra have a \neii{} HVC detection and the sensitivity is such to detect a LVC, there is no corresponding LVC \neii{} detection --- see VW~Cha and RU~Lup. We will further expand on the implications of this finding in Section~\ref{sec:discussion}.

\subsection{Comparison of line profiles}\label{sect:profiles}
Here, we compare the normalized \neii{} 12.81\,\micron{} and \oi{} 6300\,\AA{} line profiles, as well as some of the basic kinematic properties obtained by fitting them with Gaussian profiles. We discuss AA~Tau separately because of the complexity of its inner disk, multi-component \oi{} 6300\,\AA{} profile, and very large \neii{} LVC width  (Section~\ref{sec:aatau}).

For five out of eight sources, the HVC \neii{} and \oi{} profiles are similar in terms of peak centroids and overall widths (see Figure~\ref{fig:hvcprof}). Exceptions are the profiles from S~CrA, RU~Lup, and VW~Cha.
The Kendall's $\tau$ probability that the HVC centroids are not correlated with disk inclination is low (1.4\% for the \neii{} and 0.6\% for the \oi{}). As expected, the HVC centroids from both tracers become less and less blueshifted as we observe disks closer to edge-on, i.e. as  the micro-jet becomes closer to the plane of the sky and the projected radial velocity component toward the observer is reduced. No obvious trend is present between their FWHM and disk inclination (see the Kendall's $\tau$ values in the lower right panel of Figure~\ref{fig:hvcprof}). This is also expected as the line width of shocked gas traced by the HVC is mostly set by the shock kinematics, hence non Keplerian \citep[e.g.,][]{HollenbachMcKee1979,Hartigan1987}. For the \oi{} 6300\,\AA{} lines, \citet{Banzatti2019} reported the same behaviour for the HVC centroid and FWHM with disk inclination by analyzing a larger sample of disks.

\begin{figure*}[ht!]
\plottwo{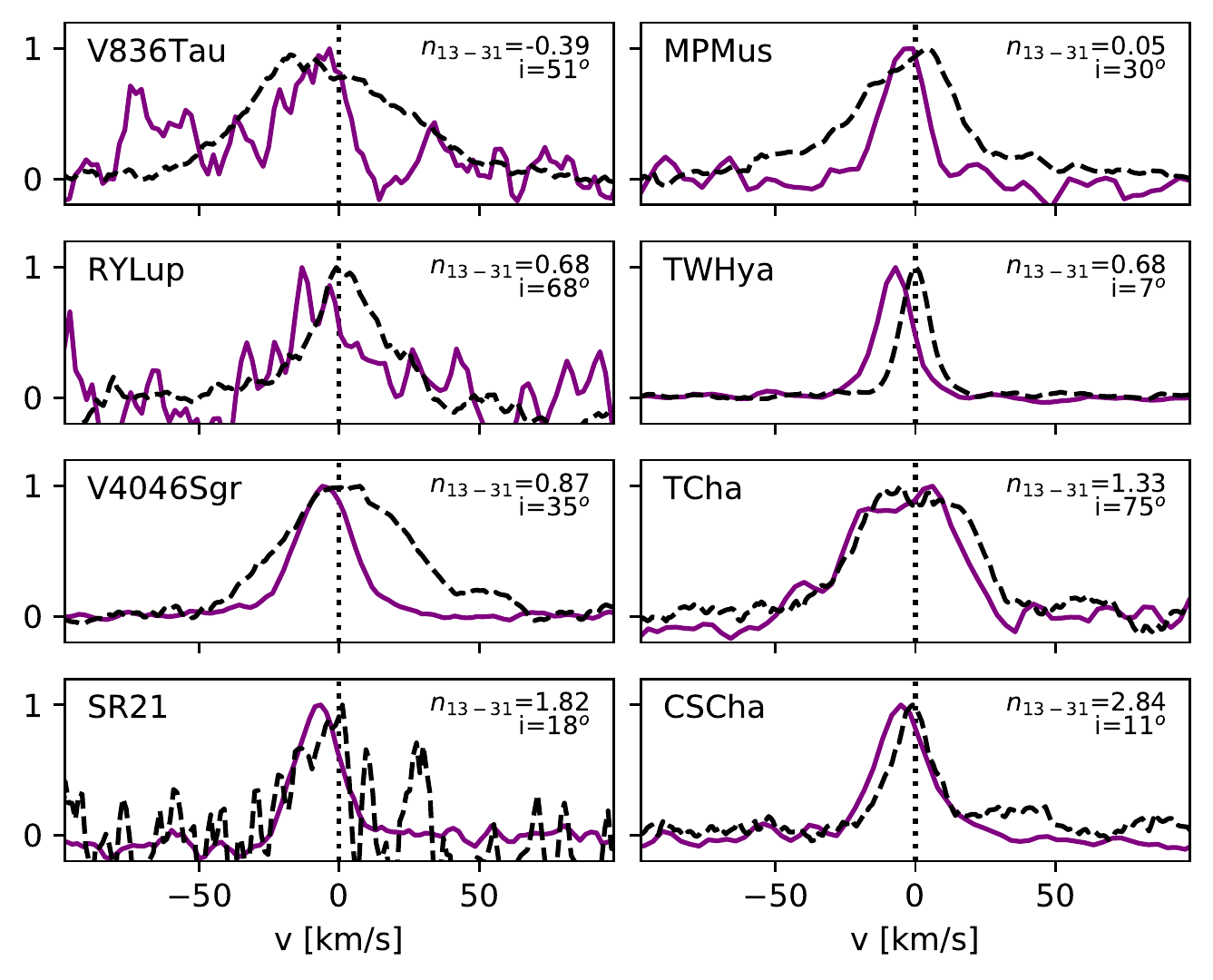}{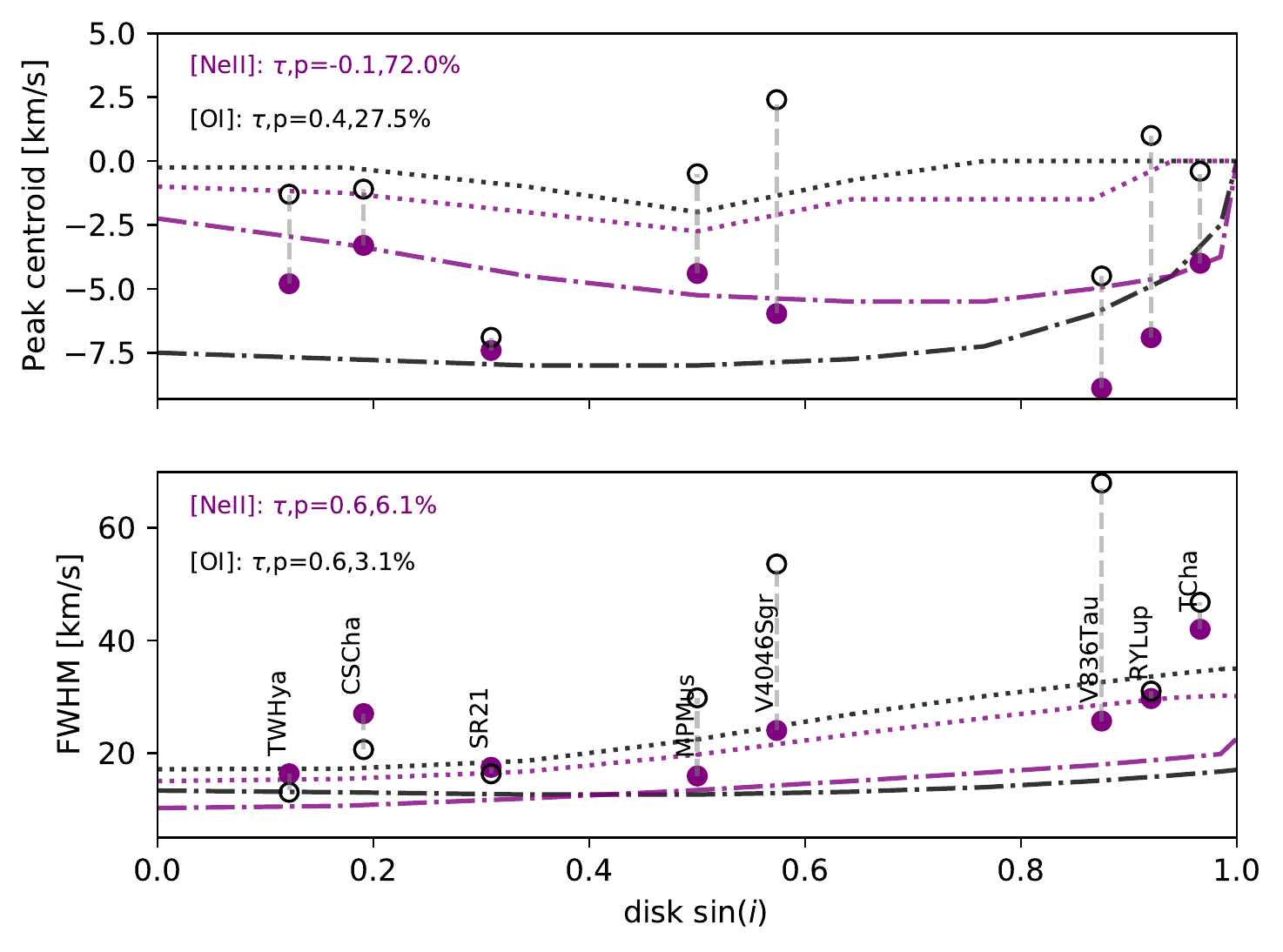}
\caption{Same as Figure~\ref{fig:hvcprof} but for the LVC, \neii{} in purple. Left panels: sources are ordered by increasing infrared spectral index ($n_{13-31}$). The disk inclination ($i$) is also provided for each target. Right panels: dot-dashed and dotted lines are  predictions from the \citet{ErcolanoOwen2010}  photoevaporative models with log($L_X$)=30.3\,erg/s for full disks and disks with a 8.3\,au hole in the gas and dust, respectively (purple lines for the \neii{} 12.81\,\micron{} while black lines for the \oi{} 6300\,\AA{} transition).
\label{fig:lvcprof}}
\end{figure*}

The LVC behavior is different: \neii{} velocity centroids are blueshifted even when the \oi{} emission is centered at the stellar velocity and the \oi{} FWHMs are larger than, or similar to, the \neii{} FWHMs (Figure~\ref{fig:lvcprof}).
There is no obvious trend between the peak centroid of both tracers and disk inclination while the LVC FWHMs tend to be larger for higher disk inclinations (see the Kendall's $\tau$ values in the right panels of Figure~\ref{fig:lvcprof}). The latter trend was already noted for the \neii{}  by \citet{Sacco2012} using VISIR~1 data while for the \oi{} it was first discussed in \citet{Simon2016} and then revised in \citet{Banzatti2019} where it is shown that  single LVC sources present the strongest correlation between the \oi{} FWHM and disk inclination. This trend suggests that Keplerian broadening plays some role in setting the  LVC line widths.
All together, this is evidence that the \neii{} 12.81\,\micron{} LVC emission traces predominantly unbound gas from a slow (small blueshifts in the peak centroids), wide-angle (lack of a correlation between centroid and disk inclination) wind. Furthermore, the correlation between FWHM and disk inclination is consistent with a photoevaporative wind or emission close enough to the base of a MHD wind for the gas to retain the Keplerian signature of the launching region. Finally, as the \neii{} LVC is always more blueshifted than the \oi{}, the wind traced by \neii{} either originates at higher elevation or at larger radii than that probed by the \oi{} 6300\,\AA{} line. To gain further insight into which of the two possibilities is the most likely, we turn to the sample of disks with inner dust depletion.

\subsubsection{Disks with dust inner cavities or large gaps}\label{sec:dustcavity}
Figure~\ref{fig:lvcprof} includes five sources whose spectral energy distribution (hereafter, SED) hinted early on at significant dust depletion in their inner disk: TW~Hya, V4046~Sgr, T~Cha, SR~21, and CS~Cha --- see \citet{vanderMa2016} for a  homogeneous analysis.
Recently, high-resolution continuum ALMA images have revealed large gaps in the case of TW~Hya ($\sim 0.5-2$\,au, \citealt{Andrews2016}), V4046~Sgr ($\sim 4-31$\,au, \citealt{Francis2020}), and T~Cha ($\sim 1-28$\,au, \citealt{Hendler2018}) while empty cavities of 56\,au, 37\,au, and 69\,au for SR~21, CS~Cha, and RY~Lup respectively \citep{Francis2020}.
 Depending on the physical process carving the cavity or gap, its radial extent could be grain-size, hence wavelength, dependent. For instance, dynamical clearing by a planet should lead to large millimeter grains
accumulating in pressure maxima beyond the planet location while small sub-micron grains are coupled with the gas and can move further in \citep[e.g.,][]{Zhu2012,deJuanOvelar2016}. This behaviour is nicely seen in the disk of T~Cha where the peak emission at 1.6\,\micron{}, tracing  sub-micron grains, is several au closer than the peak emission at ALMA wavelengths which trace millimeter grains \citep{Hendler2018}. Whether this behaviour is typical to the other disks discussed here is not known.

In relation to the \oi{} 6300\,\AA{} and \neii{} 12.81\,\micron{} line profiles, \citet{Pascucci2011} already  noted for TW~Hya that, while the former is centered at the stellar velocity, the latter is blueshifted.
Because sub-micron dust grains are the main source of opacity, the presence of these grains is necessary to shield redshifted emission  from view, to produce a blueshifted line (see photoevaporative wind models with and without cavity in \citealt{ErcolanoOwen2010}). Based on this fact, and given the decrease in dust extinction with wavelength, \citet{Pascucci2011} concluded that most of the \oi{} emission, if tracing a wind, must arise within the dust cavity while more than 80\% of the (blueshifted) \neii{} emission must arise beyond. In addition, because the small VISIR~1 slit width recovered all the {\it Spitzer}/IRS \neii{} flux, they further constrained its radial extent to within $\sim 10$\,au from the star.

 All other  disks with inner cavities or large gaps in Figure~\ref{fig:lvcprof} display the same behavior in peak centroids.
Thus, regardless of the radial extent of the cavity or gap, most of the blueshifted \neii{} emission must arise beyond the dust cavity or outside the dust gap, sometimes at tens of au, while the \oi{} emission, if tracing a wind, is radially confined within the cavity or gap.

It is also possible that the \oi{} 6300\,\AA{} line traces bound disk gas, as proposed for several Herbig Ae/Be systems \citep[e.g.,][]{Acke2005}. We follow \citet{Simon2016} in assuming a power-law distribution for the line surface brightness vs radial distance from the star  ($I_{\rm [OI]} \propto r^{-\alpha}$) and  convert it into a velocity profile assuming Keplerian rotation\footnote{routine keprot.pro by \citet{Acke2005}.}. We then convolve the model line with a velocity width that accounts for instrumental  ($\Delta v$=6.6\,km/s) and thermal broadening at 5,000\,K  as appropriate for collisionally excited gas \citep{Fang2018}. While stellar mass and disk inclination are fixed to the literature values provided in Table~\ref{tab:oiKep}, we vary the inner and outer radii of the emitting gas as well as $\alpha$ to find the best fit to the observed line profiles\footnote{using the IDL routine {\it mpfitfun} with uncertainties on the flux equal to the rms on the continuum next to the line.}. Using the best-fit
surface brightness we also compute the radius within which 90\% of the \oi{}  emission arises. As shown in Table~\ref{tab:oiKep}, the inferred power law indices for the surface brightness span only a small range between 1.5 and 2.5, in agreement with those found by \citet{Simon2016} for disks with dust cavities, and suggest that most of the emission arises close to the star, within less than a few au. Therefore, even in the case of bound disk gas, most of the \oi{} 6300\,\AA{} emission arises closer in than the \neii{} emission.

\begin{deluxetable}{lccc|ccc}
\tablecaption{Keplerian modeling for the \oi{} 6300\,\AA{} profiles of disks with inner cavities or large gaps: input and best fit parameters.}\label{tab:oiKep}
\tablewidth{0pt}
\tablehead{
\colhead{Target} & \colhead{$M_*$} & \colhead{$i$} & \colhead{$M_*$} & \colhead{$R_{\rm in}$} &   \colhead{$R_{\rm 90\%}$} & \colhead{$\alpha$}\\
\colhead{} & \colhead{(M$_\odot$)} & \colhead{($^\circ$)}  & \colhead{Ref.}  & \colhead{(au)} & \colhead{(au)} & \colhead{}
}
\startdata
RY~Lup & 1.4 & 67 & 1 & 0.29&  2.3 & 2.1 \\
TW~Hya & 0.6 & 7& 3& 0.06 &  0.4 & 2.2 \\
V4046~Sgr & 1.8 & 35 & 4 &0.27 &  1.2 & 2.5 \\
T~Cha & 1.5 & 75 & 5 & 0.15 & 2.9 & 1.5 \\
SR~21 & 1.8 & 18 & 3 & 0.08 &  0.8 & 1.9 \\
CS~Cha & 1.3 & 11 & 6& 0.02 & 0.17 & 2.1 \\
\enddata
\tablerefs{1. \citet{Hendler2020}; 3. \citet{Fang2018}; 4. \citet{Rosenfeld2013}; 5. \citet{Schisano2009}; 6. \citet{Pascucci2016}. References for disk inclinations are given in Table~\ref{tab:neii}.}
\end{deluxetable}

\subsubsection{AA~Tau}\label{sec:aatau}
AA~Tau has been long known to have a peculiar light curve with quasi-cyclic fading episodes at optical wavelengths interpreted as periodic occultations of the star by a warped inner disk \citep[e.g.,][]{Bouvier1999}. While the inner disk is thought to be viewed close to edge-on, the outer disk was recently imaged by ALMA and found to be only modestly inclined \citep[59$^\circ$,][]{Loomis2017}.

\begin{figure}[ht!]
\plotone{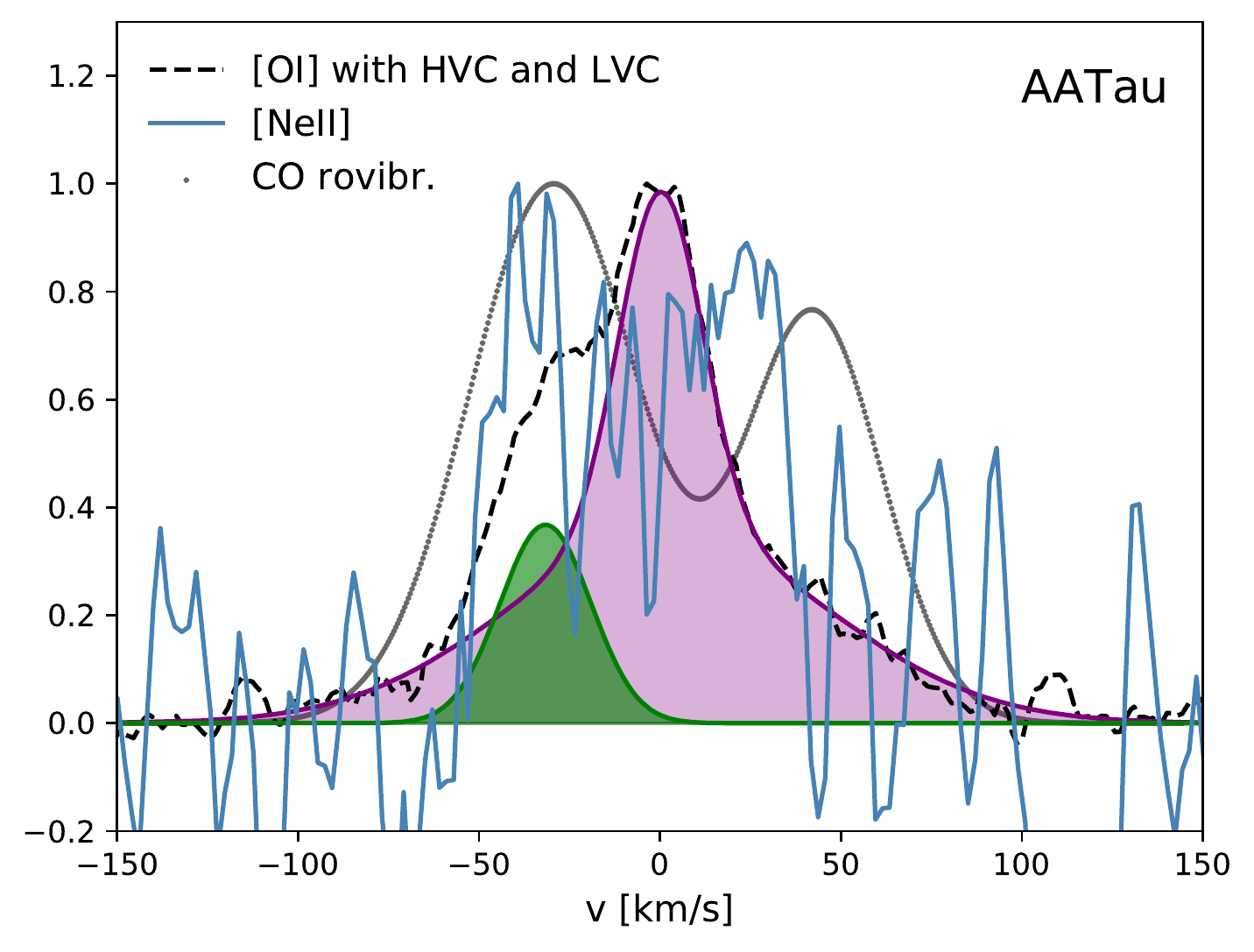}
\caption{Comparison of line profiles for AA~Tau in the stellocentric reference frame: \oi{} 6300\,\AA{} (black dashed line) with HVC (green) and LVC (purple) decomposition \citep{Banzatti2019}; \neii{} 12.81\,\micron{} emission (blue solid line, this work); stacked CO rovibrational profile at 4.7\,\micron{} \citep[grey dots,][]{BanzattiPontoppidan2015}. \label{fig:aatau}}
\end{figure}

As shown in Figure~\ref{fig:aatau}, the \oi{} 6300\,\AA{} profile from AA~Tau is complex: the HVC (green) is blueshifted by only $\sim 30$\,km/s, probably due to the viewing angle, while the LVC (purple) is characterized by a narrow peak and broad wings \citep{Simon2016,Banzatti2019}. The \neii{} profile (solid blue), although of much lower S/N, is clearly broader (FWHM$\sim 84$\,km/s) than the \oi{} HVC (FWHM$\sim 30$\,km/s), it is not centrally peaked as the \oi{} LVC but still fairly symmetric around the stellar velocity. While contamination from HVC emission is possible, the properties of the \neii{} profile hint at bound disk gas as the dominant emitting region. The same conclusion was reached by \citet{najita2009} who recovered a broad \neii{} FWHM of $\sim 70$\,km/s in spite of technical issues in retrieving the blue portion of their  TEXES spectrum ($R \sim 80,000$). For comparison, we also show in Figure~\ref{fig:aatau}
the  normalized stacked 4.7\,\micron{} CO ro-vibrational profile (grey) with the two peaks characteristic of emission from a Keplerian disk. The CO profile  is even broader (FWHM=115\,km/s) than the \neii{} pointing to a CO emitting radius of only 0.2\,au \citep{BanzattiPontoppidan2015}.
If \neii{} at 12.81\,\micron{} also traces bound disk gas, it should probe a larger range of disk radii further out than the CO. A higher S/N spectrum is necessary to better constrain the \neii{} emitting region from AA~Tau.

\section{Discussion} \label{sec:discussion}
By analyzing a large sample of high-resolution ($\Delta v \sim $10\,km/s) mid-infrared spectra, we found that the forbidden \neii{} line at 12.81\,\micron{}, similarly to the \oi{} 6300\,\AA{} line, mostly traces unbound gas flowing away from the star+disk system (see the line centroids v$_{\rm c}$ in Table~\ref{tab:neii}).
Following the kinematic classification applied to optical forbidden lines and by comparing line profiles and basic kinematic properties, we have also evidence for the \neii{} HVC to originate in fast collimated micro-jets while the LVC might trace a slower disk wind.

However, there are also important differences between these two tracers. First, while the \oi{} HVCs are typically accompanied by LVCs \citep{Banzatti2019}, the \neii{} detections show either a HVC or a LVC, with about an equal number of the two components in these spectra (Table~\ref{tab:neii}). Second, while the $L_{\rm [OI]_{LVC}}$ decreases as the dust inner disk is depleted (higher n$_{\rm 13-31}$ index), the $L_{\rm [NeII]_{LVC}}$ increases (Figure~\ref{fig:oineii} bottom panels). Finally, while most HVCs have similar morphologies in the two forbidden lines (Figure~\ref{fig:hvcprof}), the \neii{} LVC profiles, most of which are for  systems with dust depleted inner disks, are typically more blueshifted than the \oi{} LVC (Figure~\ref{fig:lvcprof}). Interestingly, even disks with tens of au mm dust cavities like SR~21 or CS~Cha present blueshifted \neii{} emission pointing to slow winds outside the gravitational radius \citep[e.g., eq.~2 in][]{Alexander2014}. In contrast, their \oi{} emission is centered at the stellar radial velocity and could thus arise from  bound disk gas (Table~\ref{tab:oiKep}) or a wind inside the dust cavity, as already discussed for TW~Hya \citep{Pascucci2011}. Thus, by combining the \oi{} and \neii{} diagnostics, we find evidence of slow flows, possibly disk winds, at all disk evolutionary stages.

In the following, we will discuss the ionization of Ne atoms and how our observations compare with predictions from static disk and wind models ( Sections~\ref{sec:ionization} and \ref{sec:comparison}).  As none of the current models can fit the entirety of the data at hand, we also sketch an evolutionary scenario that might explain the existing data (Section~\ref{sec:scenario}).

\subsection{Static disk models and ionization of Ne atoms}\label{sec:ionization}
\citet{HollenbachGorti2009}  pointed out that, if one considers reasonable EUV and X-ray spectra for young stars, the X-ray-heated disk layer produces more \neii{} 12.81\,\micron{} emission than the EUV-heated layer because in the latter most photons are used to ionize H rather than Ne (see also \citealt{Glassgold2007,ErcolanoOwen2010,Aresu2012}). Observations also suggest that the \neii{} emission mostly traces an X-ray rather than an EUV layer. If the EUV luminosity scales with $\nu^{-1}$, as suggested by the fact that $\nu L_{\nu}$ in the FUV is about the same as in the X-ray, the \neiii{} 15.5\,\micron{} luminosity should be higher than the \neii{} at 12.81\,\micron{} in the EUV-heated layer \citep[Fig.~1 in][]{HollenbachGorti2009}. However, the \neiii{} 15.5\,\micron{} line is rarely detected in {\it Spitzer}/IRS spectra and, when detected, the \neiii{}-to-\neii{} line flux ratios are significantly less than 1, except for SZ~Cha where the ratio is close to unity \citep[e.g.,][]{Najita2010,Szulagyi2012,Espaillat2013}. Regardless of the EUV spectrum, centimetric radio data  demonstrate that the EUV luminosity impinging on the disk surface is too low to reproduce the observed \neii{} 12.81\,\micron{} luminosities \citep{Pascucci2014}. This means that Ne atoms are ionized by 1\,keV hard X-ray photons via the Auger effect. X-rays are also likely to heat the gas in the same region, although  it remains unclear whether they are the dominant heating source.

A  prediction from static disk models is that the \neii{} 12.81\,\micron{} luminosity should scale almost linearly with $L_X$ in the X-ray-heated disk layer \citep[e.g.,][]{HollenbachGorti2009,Aresu2012}. While several of the data points fall on the predicted relation (see Figure~\ref{fig:oineii}), the current dataset cannot confirm the existence of a correlation between $L_{\rm [NeII]}$ and $L_X$. On the one hand, it would be important to expand the sample to cover a broader range of \neii{} and $L_X$ luminosities. On the other hand, it would be interesting to test if the predicted $L_{\rm [NeII]} - L_X$ correlation for a static atmosphere holds also in a flowing one \citep[e.g.,][]{ErcolanoOwen2010} as the data demonstrate that the \neii{} 12.81\,\micron{} line mostly traces unbound gas (Section~\ref{sect:profiles}).

\subsection{Photoevaporative and MHD disk wind models}\label{sec:comparison}
As summarized in the Introduction, significant progress has been made recently in understanding  the origin of optical forbidden lines, such as the \oi{} 6300\,\AA{}. In particular, there is consensus on its HVC primarily tracing a micro-jet and its LVC-BC tracing an inner MHD wind \citep[e.g.,][]{Ray2007,Simon2016,McGinnis2018,Weber2020}. Uncertainty remains on the LVC-NC which is either attributed to some other region in the inner MHD wind  \citep[e.g.,][]{Banzatti2019} or to an outer photoevaporative wind \citep[e.g.,][]{Weber2020}.

For the \neii{} 12.81\,\micron{} line investigated here, the spectral resolution and sensitivity do not allow to further decompose identified LVCs into BC and NC. Hence, we can only treat the entire \neii{} LVC as a single phenomenon. We will examine our results in the context of  photoevaporative models and then move on to  recent analytic MHD disk wind models.

In relation to photoevaporative X-EUV models, \citet{ErcolanoOwen2010} find that disks with an inner hole in the dust and gas produce a factor of $\sim 2$ higher $L_{\rm [NeII]}$ than full disks, mostly because stellar X-rays are not absorbed by the inner disk and can thus heat and ionize a larger/further away portion of the wind. This trend agrees with our finding that the $L_{\rm [NeII]_{LVC}}$ is higher for sources with dust depleted inner disks (Figure~\ref{fig:oineii} bottom right panel).
However, the predicted \neii{}-to-\oi{} line ratios are rather similar (grey horizontal band in Figure~\ref{fig:neii_oi}) and cannot explain the  majority of our data, including the many stringent LVC upper limits (downward triangles).
In addition,  because the predicted \oi{} 6300\,\AA{} line is extremely sensitive to high temperatures (E$_{\rm up}\sim 22,830$\,K), it probes the hot inner portion of the wind closer to the star than the \neii{} line, hence the \oi{} profiles are more blueshifted than the \neii{} for full disks, or both mostly at the stellar velocity for disks with holes (see Figure~\ref{fig:lvcprof} as well as \citealt{Picogna2019}). Both of these trends are opposite to what is observed.

 \citet{Ballabio2020} used the self-similar solutions for thermal disk winds developed by \citet{ClarkeAlexander2016} to calculate \oi{} and \neii{} line profiles for different disk inclinations.
 For a given sound speed and the same disk inclination, the models predict similar blueshifts for the \neii{} and \oi{} lines while the \oi{} FWHM tends to be larger than the \neii{}. When compared with previous data drawn from the literature, the model for a 10\,km/s thermal wind successfully reproduces the blueshifts and widths of the observed \neii{} LVC lines.
 The observed \oi{} line blueshifts favour cooler low-velocity gas models ($c_{\mathrm s}=3-5$~km/s), but the predicted widths for these models are less than 15\,km/s, much smaller than the observed ones. This suggests that a single thermal wind model can not easily reproduce the observations of both \neii{} and \oi{} 6300\,\AA{} LVC lines.

Recently, \citet{Weber2020} computed line profiles from an X-ray photoevaporative wind model based on the radiation-hydrodynamical
calculations of \citet{Picogna2019} and an analytic MHD wind model following \citet{BP82} with photoionization as in \citet{ErcolanoOwen2010}. Although they do not discuss the \neii{} 12.81\,\micron{} emission, they cover several optical forbidden lines, including the \oi{} 6300\,\AA .
As in their MHD models the high critical density  \sii{} 4068, \oi{} 5577 \& \oi{} 6300\,\AA{} lines  come from within a few au and close to the disk surface, the synthetic profiles have Keplerian double peaks\footnote{Note that line profiles from MHD wind models are very sensitive to the assumed, and not well constrained, structure of the flow (see e.g., the different forbidden line profiles in \citealt{Garcia2001,Shang2010})}  for disks inclined by more than $30^\circ$, which are not observed \citep[e.g.,][]{Banzatti2019}.
They conclude that if the MHD wind (producing the HVC and BC) is also accompanied by a photoevaporative wind,
then the Keplerian trough in the BC can be filled by the narrow component emission from the thermal wind.
Our major concern with the addition of such a photoevaporative wind is in its radially and vertically extended lower density region that is  bright in the \sii{} 6730\,\AA{} line (Fig.~1 in \citealt{Weber2020}).
The predicted \sii{} 6730\,\AA{} line luminosities
are similar to, or higher than, the \oi{} 6300\,\AA{} luminosities and the mean \sii{} 6730 over 4068\,\AA{} line ratio for models with accretion luminosity comparable to the observed ones ($L_{\rm acc}=3\times10^{-2}$\,L$_\odot$) is slightly above unity.
On the contrary observations show that the \sii{} 6730\,\AA{} line is far less common than the \oi{} 6300\,\AA{} or the \sii{} 4068\,\AA{}, when detected it is usually as an HVC, not an LVC, and, combining contemporaneous LVC detections and non-detections, the mean  \sii{} 6730 over 4068\,\AA{} line ratio is $\sim 0.15$, well below unity \citep[e.g.,][]{Hartigan1995,Pascucci2011,Natta2014,Simon2016,Fang2018}.

As MHD wind models have a less extended low density region \citep{Weber2020} and considering the empirical correlations between the \oi{} HVC and its LVB-BC and NC \citep{Banzatti2019}, we lean towards prior interpretations that
attribute the optical forbidden lines solely to an MHD wind and a jet in  systems with full disks and high accretion rates \citep[e.g.,][]{Natta2014,Fang2018,Nisini2018,Banzatti2019}.

\subsection{ An evolutionary sketch for disk winds} \label{sec:scenario}
Given that none of the current models exactly fits the observations at hand (Section~\ref{sec:comparison}), we put forth an empirically motivated evolutionary scenario that can be tested by future observations and disk wind models.

One of the most important results from our study is that the majority (9/11) of full disks, i.e. disks with $n_{\rm 13-31} < 0$, have an inner MHD disk wind identified via the \oi{} 6300\,\AA{} LVC emission but lack a \neii{} 12.8\,\micron{} LVC detection (Figure~\ref{fig:neii_oi}).
As shown in Figure~\ref{fig:neii_oi} and Table~\ref{tab:neii}, 3$\sigma$ $L_{\rm [NeII]}$ upper limits, calculated for a FWHM that is appropriate for a LVC, are stringent enough for most sources and suggest that the \neii{} LVC in full disks is several times weaker than in  disks with inner dust depletion while the reverse is true for \oi{} 6300\,\AA{} (which emission diminishes in disks with larger dust cavities).  These results strongly suggest that  hard X-ray photons ionizing Ne atoms are somehow screened in full disks and do not pass beyond the inner wind where they would produce enough detectable \neii{} emission as shown e.g. in the fully atomic photoevaporative models of \cite{ErcolanoOwen2010}, see their Fig.~3.

\begin{figure}[ht!]
\plotone{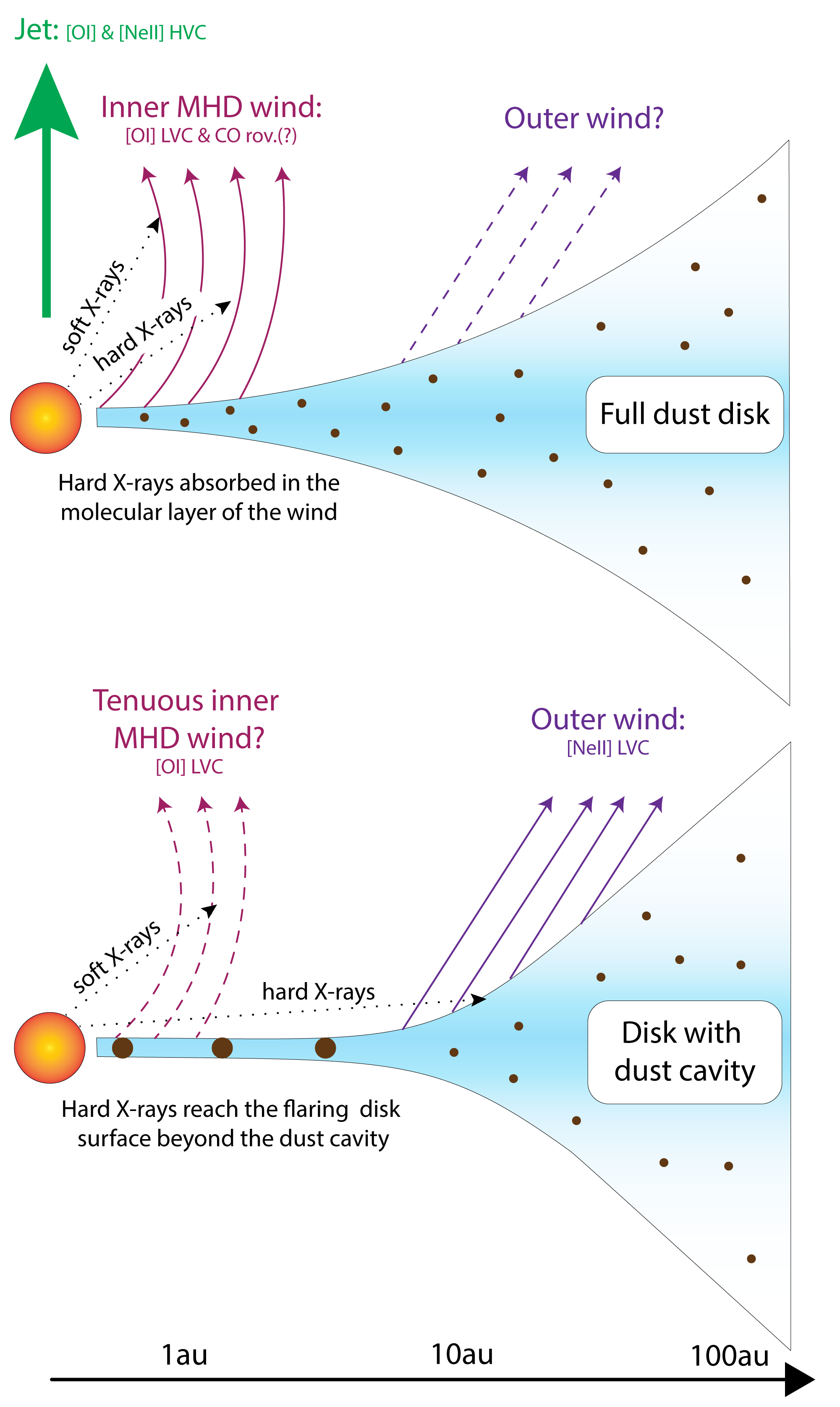}
\caption{Evolutionary sketch. Upper figure: full dust disk with an inner MHD wind that screens X-ray photons. Lower figure: disk with dust cavity and a tenuous inner wind that enables hard X-ray photons to penetrate deeper and produce detectable \neii{} LVC emission. The source in the upper panel is a higher accretor than the one on the lower panel and powers a jet detected as HVC. \label{fig:diskwind}}
\end{figure}

\citet{HollenbachGorti2009} discussed the penetration of high-energy stellar photons through inner winds, which they modelled following the ``X wind'' prescription \citep{Shu1994}. As such, these inner winds arise within 10 stellar radii where all dust has likely sublimated. They find that a gas column density of $\sim 10^{22}$\,cm$^{-2}$ is required for 1\,keV optical depth of unity  which translates into a wind mass loss rate less than $\sim 4 \times 10^{-8}$\,M$_\odot$/yr for hard X-rays to penetrate the wind. Interestingly, this value is within a factor of a few of the mass accretion rate below which \neii{} LVC detections dominate (Figures~\ref{fig:histoNeII} and \ref{fig:oineii}).
Soft X-rays peaking at $\sim 0.2$\,keV are screened by a column of only $\sim 10^{20}$\,cm$^{-2}$, hence they are mostly absorbed at the surface of the wind exposed to the star: this behavior is seen in the models of \cite{ErcolanoOwen2010} through the \oi{} 6300\,\AA{} line which mostly traces soft X-rays.
But different penetration depths alone are unlikely to explain the ensemble of the observations. The excitation temperature of the \neii{} line is only $\sim$1100\,K, and therefore the gas would have to be cooler than a few 100\,K to not excite the transition, fully atomic gas cannot typically cool efficiently \citep[e.g.,][]{Ercolano2008}. However, if the portion of the inner wind where hard X-rays are absorbed is mostly molecular, efficient cooling  \citep[e.g.,][]{Gorti2016} could suppress the \neii{} emission below detectable values.

Figure~\ref{fig:diskwind} sketches a possible evolutionary scenario.
The upper panel illustrates a typical full, flared disk with an inner MHD wind that is mostly atomic out to a radial distance where soft-Xrays penetrate. This hot ($>$\,5,000\,K) atomic layer, perhaps heated also by ambipolar diffusion \citep[e.g.,][]{Safier1993}, would be responsible for the \oi{} 6300\,\AA{} LVC. Hard X-rays penetrate deeper in the wind and  would be absorbed in a mostly molecular layer that is too cool to produce  detectable \neii{} emission. The narrow component  (FWHM$= 10–50$\,km/s) of the CO fundamental emission is found to trace gas of $200-700$\,K \citep[e.g.,][]{BanzattiPontoppidan2015}, hence might trace this cooler molecular wind. Interestingly, RU~Lup and S~CrA, two of our disks that have \oi{} HVC \& LVC emission but only a \neii{} HVC component (see Figure~\ref{fig:hvcprof}), present astrometric signals in their CO fundamental lines that are consistent with a molecular wind \citep{Pontoppidan2011}.

The lower panel of Figure~\ref{fig:diskwind}  illustrates a typical disk with a dust inner cavity, lower accretion and weaker inner wind. At this stage X-ray photons penetrate deeper. The \oi{} LVC emission, which is correlated with mass accretion rate (Figure~\ref{fig:oineii}), weakens and could trace larger radii, as suggested by the positive correlation between FWHM and spectral index \citep{Banzatti2019}. Still, most of the emission would arise within the dust cavity (Section~\ref{sec:dustcavity}).  The molecular layer is also significantly reduced which is supported by the finding that  nearly all  disks with dust depleted inner cavities show no evidence for infrared water emission lines in their {\it Spitzer}/IRS spectra and CO emitting radii are larger \citep{Salyk2015,Banzatti2017}. The hard 1\,keV X-ray photons would pass through the inner wind and reach the edge of the dust cavity where the disk starts flaring: there, they would ionize Ne atoms and produce detectable \neii{} 12.81\,\micron{} emission in an outer wind. As discussed in Section~\ref{sect:profiles}, the lack of correlation between \neii{} LVC centroid and disk inclination suggests that this outer wind has a wide opening angle. However, current data are not sufficient to establish whether the wind is photoevaporative or MHD  in nature and its presence does not imply that the dust cavity is opened by the wind.

Recently \citet{Simon2018} put forward a preliminary model to reconcile largely laminar MHD winds, which produce significant turbulent velocities in the outer disk, and the limits on turbulent broadening obtained with ALMA. A key component of this model is a massive wind inside 30\,au that would block high-energy stellar photons, in particular X-rays and FUV photons.
While our data do not constrain the screening of FUV photons, they provide {\bf some} evidence for the inner region of full disks blocking X-rays and not reaching the outer disk.

\section{Summary}
We have analyzed a sample of 31 disks that were observed with high-resolution optical ($\Delta v \sim 7$\,km/s) and infrared ($\Delta v \sim 10$\,km/s) spectra covering the \oi{} 6300\,\AA{} and the \neii{} 12.81\,\micron{} lines. Our VISIR~2 infrared survey discovered 6 new \neii{} detections and confirmed 5  detections previously reported in the literature. Following analysis carried out at optical wavelengths \citep[e.g.,][]{Simon2016}, we fit the detected lines with Gaussian profiles and classified them into HVC (or LVC) depending on whether the line centroid is more (or less) blueshifted than 30\,km/s from the stellar radial velocity. Combining the VISIR~2 and literature VISIR~1 results, we explored if the known correlations between the \oi{} luminosities and stellar/disk properties apply to the \neii{}  luminosities. In addition, we compared the detected \neii{} 12.81\,\micron{} and \oi{} 6300\,\AA{} profiles  to investigate whether these transitions trace the same region.
Our main results can be summarized as follows:
\begin{itemize}
    \item All 17 sources with \neii{} detections present either a HVC or a LVC, in about equal number. This is very different from the \oi{} 6300\,\AA{}, where the LVC is found in most sources, many of which also have a HVC  \citep{Banzatti2019}.
    \item High accretors ($\dot{M}_{\rm acc} > 10^{-8}$\,M$_\odot$/yr) with \neii{} detections present only a HVC in this line. \neii{} LVCs are preferentially detected in sources with low $\dot{M}_{\rm acc}$, low \oi{} 6300\,\AA{} emission, and high infrared spectral index ($n_{\rm 13-31}$).
    \item The \neii{} and \oi{} luminosities display the opposite behavior with $n_{\rm 13-31}$:  while the $L_{\rm [OI]_{LVC}}$ decreases as the dust inner disk is depleted (higher $n_{\rm 13-31}$ index), the $L_{\rm [NeII]_{LVC}}$ increases.
    \item The \neii{} and \oi{} HVC profiles are generally similar with centroids and FWHMs showing the expected behaviour from shocked gas in micro-jets, i.e. centroids less blueshifted for closer to edge-on disks and FWHMs independent of disk inclination.
    \item The \neii{} LVC profiles are more blueshifted and typically narrower than the \oi{} LVC profiles. Both FWHMs increase for closer to edge-on disks while the peak centroids are independent of disk inclination. This is evidence that the \neii{} LVC predominantly traces unbound gas from a slow, wide-angle wind that has not completely lost the Keplerian signature from its launching region.
\end{itemize}
As the majority of full disks in our survey have an inner MHD disk wind detected via the \oi{} LVC but lack a \neii{} LVC, we infer that the $\sim 1$\,keV hard X-ray photons needed to ionize Ne atoms are somehow screened in full disks. We  suggest that this screening occurs in inner MHD disk winds which, except for a hot surface exposed to the star and likely traced via the \oi{} LVC, are mostly molecular. The molecular wind, suggested in a few sources via CO spectroastrometry \citep{Pontoppidan2011},  would absorb most of the hard X-rays and, thanks to efficient cooling, produce little \neii{} emission, below the VISIR sensitivity. As the dust inner disk is depleted and the inner wind and accretion weaken,
 hard X-rays can penetrate deeper reaching the outer edge of the dust cavity where the gas disk flares. There, they would heat and ionize a larger surface area which becomes detectable in our survey via the \neii{} 12.81\,\micron{} line. At the same time, the \oi{} 6300\,\AA{} luminosity, which strongly correlates with $\dot{M}_{\rm acc}$, diminishes and, being confined within the dust cavity, can lose the kinematic signature of a wind. JWST/MIRI observations of selected disks with large dust cavities, like SR~21 and RY~Lup, could spatially resolve the outer wind in the \neii{} 12.81\,\micron{} line.

Recently, \citet{Fang2018} used a combination of optical forbidden lines from atomic and ionic species to estimate the LVC wind mass loss rates. If our evolutionary sketch is correct and the inner MHD winds from full disks have a significant molecular component, the \citet{Fang2018} values should be taken as lower limits. Expanding the set of molecular diagnostics for the inner winds, as well as the sample of disks with detections, to date restricted to suggestive evidence only in a handful of disks, will be important to better constrain wind mass loss rates.
Disk wind models that properly couple thermodynamics and hydrodynamics  are being developed \citep[e.g.,][]{Gressel2020} and are necessary to test our proposed evolutionary scenario.

\acknowledgments
Based on observations collected at the European Southern Observatory under ESO programme 198.C-0104.
I.P., U.G., and S.E. acknowledge support from a Collaborative NSF Astronomy \& Astrophysics Research grant (ID: 1715022, ID:1713780, and
ID:1714229).
This
material is based on work supported by the National
Aeronautics and Space Administration under Agreement No.
NNX15AD94G for the program {\it Earths in Other Solar Systems}.
The results reported herein benefitted from collaborations and/or information exchange within NASA’s Nexus for Exoplanet
System Science (NExSS) research coordination network
sponsored by NASA’s Science Mission Directorate.
R.D.A., G.B., and C.H. acknowledge funding from the European Research Council (ERC) under the European Union's Horizon 2020 research and innovation program (grant agreement No 681601). G.B. acknowledges support from the University of Leicester through a College of Science and Engineering PhD studentship. C.H. is a Winton Fellow and this research has been supported by Winton Philanthropies / The David and Claudia Harding Foundation. This project has been carried out as part of the European Union's Horizon 2020 research and innovation programme under the Marie Sk\l odowska-Curie grant agreement No 823823 ({\sc dustbusters}).

\facilities{ESO(VISIR)}

\software{{\tt astropy} (The Astropy Collaboration 2013, 2018), {\tt mpfitfun} (MINPACK-1 in Jorge More' \& Stephen Wright 1987), {\tt cenken} (Helsel 2005 \& Akritas, Murphy, and LaValley 1995)}


\newpage
\appendix
\section{Reduction of unpublished optical spectra and \oi{} line luminosities}\label{app:archive}
The sources CS~Cha, MP~Mus, T~Cha, TY~CrA, T~CrA, V4046~Sgr, and VW~Cha have very noisy or no published high-resolution optical spectra. For all of them except TY~CrA,  we retrieved from the ESO archive\footnote{http://archive.eso.org/cms.html} the highest signal-to-noise UVES  spectra (R$\sim$45,000) available\footnote{In the case V4046~Sgr we combine six UVES spectra to increase the signal-to-noise. }. For TY~CrA we used an unpublished Magellan/MIKE spectrum at a similar spectral resolution (R$\sim$42,000). The data reduction was carried out as in \citet{Fang2018} and \citet{Banzatti2019} and includes: removal of  telluric absorption lines and photospheric features as well as subtraction of the stellar radial velocity to bring the spectra in the stellocentric reference frame.
The stellar radial velocities  for all sources except TY~CrA and V4046~Sgr are derived by cross-correlating each star optical spectrum with the synthetic spectrum of a star that has the same effective temperature (see Table~\ref{tab:targets}). For TY~CrA and V4046~Sgr we use the center of mass radial velocity of the eclipsing binary and that of the circumbinary CO disk, respectively, further details in Appendix~\ref{app:notes}.
Figure~\ref{fig:newoi} shows the seven archival spectra around the \oi{} 6300\,\AA{} line in the stellocentric reference frame.  The noisy continuum of V4046~Sgr between -300 and -100\,km/s is due to the challenge in properly subtracting the photospheric absorption from this close binary system.

\begin{figure}[h!]
\plotone{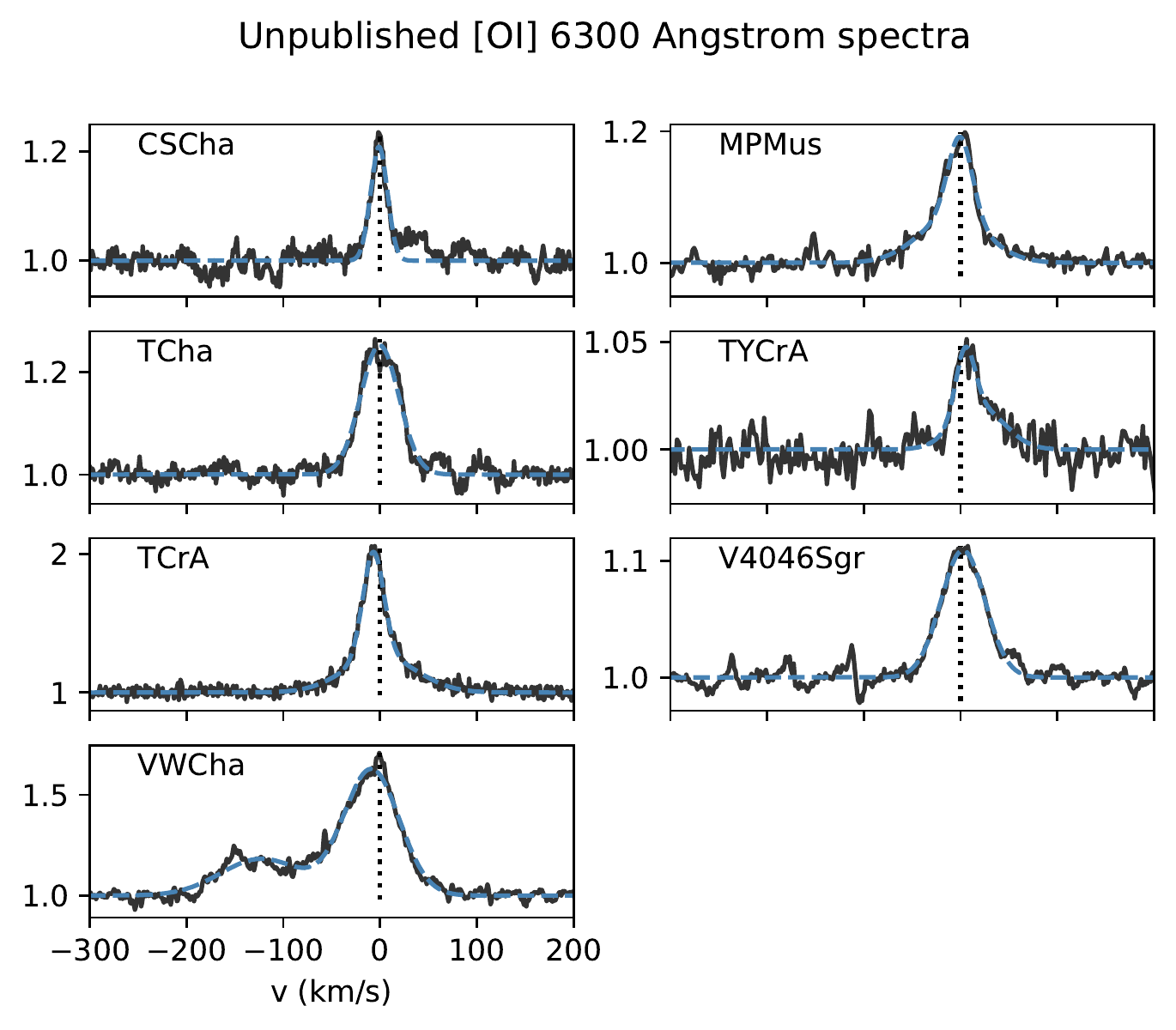}
\caption{Unpublished VLT/UVES and MIKE/Magellan spectra utilized in this work showing the \oi{} 6300\,\AA{} emission in the stellocentric reference frame. The blue dashed line in each panel shows the best fit using the Gaussian components in Table~\ref{tab:oinew}. \label{fig:newoi}}
\end{figure}

Following \citet{Simon2016} we fit the minimum number of Gaussian profiles to reproduce the observed lines: first we fit just one Gaussian, then we compute the rms of the spectrum minus Gaussian at the location of the emission, we add another Gaussian if this rms is larger than twice the rms outside the line. With this approach we find that the spectra of CS~Cha\footnote{After experimenting with different standard stars, we conclude that the extra emission on the red side of the \oi{} line is likely due to poor photospheric correction, hence we do not fit it.}, T~Cha, and V4046~Sgr can be fit by just one Gaussian profile while those of MP~Mus, TY~CrA, T~CrA, and VW~Cha require two Gaussian profiles. For each component, the FWHM, Gaussian centroid (v$_{\rm c}$), and EW are provided inTable~\ref{tab:oinew} together with a classification (Type) that follows \citet{Simon2016}. Based on this classification, a component is called LVC (HVC) if the absolute value of the Gaussian velocity centroid is smaller (larger) than 30\,km/s and, within the LVC, a BC has a $FWHM > 40$\,km/s while the NC is narrower. Only VW~Cha in this sample has a clearly distinct high-velocity component (HVC) associated with a jet while the viewing angle of TY~CrA and T~CrA is such that most of the \oi{} emission is likely from a jet (HVC), see Appendix~\ref{app:notes}.

To convert the measured EWs into line luminosities we collect V magnitudes and extinctions (A$_{V}$) from the literature, use the source distances in Table~\ref{tab:targets},  and take the flux of a zero magnitude star in the V band from \citet{Bessell1979}, $F_V = 3.1\time 10^{-9}$ erg/cm$^2$/s/\AA{}. With these inputs the \oi{} 6300\,\AA{} luminosity is calculated as:
\begin{equation}
 L_{\rm [OI]} = 10^{-(V-Av)/2.5} \times F_V \times EW \times (4 \pi {\rm Dist}^2)
\end{equation}
When multiple kinematic components are identified, we provide in Table~\ref{tab:targets} the total \oi{} luminosity obtained from the sum of all components while the LVC contains the BC and NC contributions.

\begin{deluxetable}{lcccccccc}
\tablecaption{\oi{} 6300\,\AA{} decomposition from unpublished high-resolution spectra. \label{tab:oinew}}
\tablewidth{0pt}
\tablehead{
\colhead{Target} & \colhead{FWHM} & \colhead{v$_{\rm c}$} & \colhead{EW}  & \colhead{Type} & \colhead{V} & \colhead{A$_{\rm V}$} & \colhead{V,A$_{\rm V}$} &\colhead{Log $L_{\rm [OI]}$}\\
\colhead{} & \colhead{(km/s)} & \colhead{(km/s)} & \colhead{(\AA)} & \colhead{} & \colhead{}  & \colhead{}  & \colhead{Ref.} &\colhead{($L_\odot$)}
}
\startdata
CSCha & 20.6 & -1.1 & 0.1 & LVC-NC & 11.7 & 0.8 & 1 & -4.9 \\
MPMus & 29.8 & -0.5 & 0.08 & LVC-NC & 10.44 & 0.17 & 2 &-5.21 \\
            & 82.8 & -8.6 & 0.13 & LVC-BC &   &   & & -5.04 \\
TCha & 46.8 & -0.4 & 0.26 & LVC-BC & 11.6 & 1.3 & 3 & -4.64 \\
TYCrA & 23.2 & 4.6 & 0.02 & LVC$\rightarrow$HVC\tablenotemark{\dag} & 9.26 & 1.98 & 4 & -4.44 \\
  & 61.8 & 19.3 & 0.03 & LVC$\rightarrow$HVC\tablenotemark{\dag} &  &  & & -4.23 \\
TCrA & 25.4 & -7.0 & 0.43 & LVC$\rightarrow$HVC\tablenotemark{\dag} & 12.04 & 1.9 & 5 & -4.06 \\
  & 87.9 & 0.4 & 0.51 & LVC$\rightarrow$HVC\tablenotemark{\dag} &   &   & & -3.99 \\
V4046Sgr & 53.6 & 2.4 & 0.13 & LVC-BC & 10.68 & 0.0 & 6 & -5.45 \\
VWCha & 67.7 & -8.8 & 0.95 & LVC-BC & 12.8 & 1.9 & 7 & -3.84 \\
  & 98.8 & -122.6 & 0.4 & HVC &   &   & & -4.21 \\
\enddata
\tablecomments{The V magnitude of T~Cha varies by up to 1.5\,mag, we report here its median value.}
\tablenotetext{\dag}{The 2D spectrum of TCrA reveals a jet close to the plane of the sky, hence it is not possible to distinguish the jet emission from any LVC (Whelan et al. in prep.). The disk around the quartenary star system TY~CrA is likely seen edge-on, hence any jet emission will be close to the plane of the sky. Our assignment is corroborated by the \sii{} 6731\,\AA{} detection which is sensitive to jet emission and peaks very close to the \oi{} 6300\,\AA{} lowest velocity component.}
\tablerefs{1. \citet{Espaillat2007}, 2. \citet{Mamajek2002}, 3. \citet{Schisano2009}, 4. \citet{Vioque2018}, 5. \citet{GarciaLopez2006}, 6. \citet{Donati2011}, 7. \citet{Manara2017} }
\end{deluxetable}

\begin{figure}[ht!]
\plotone{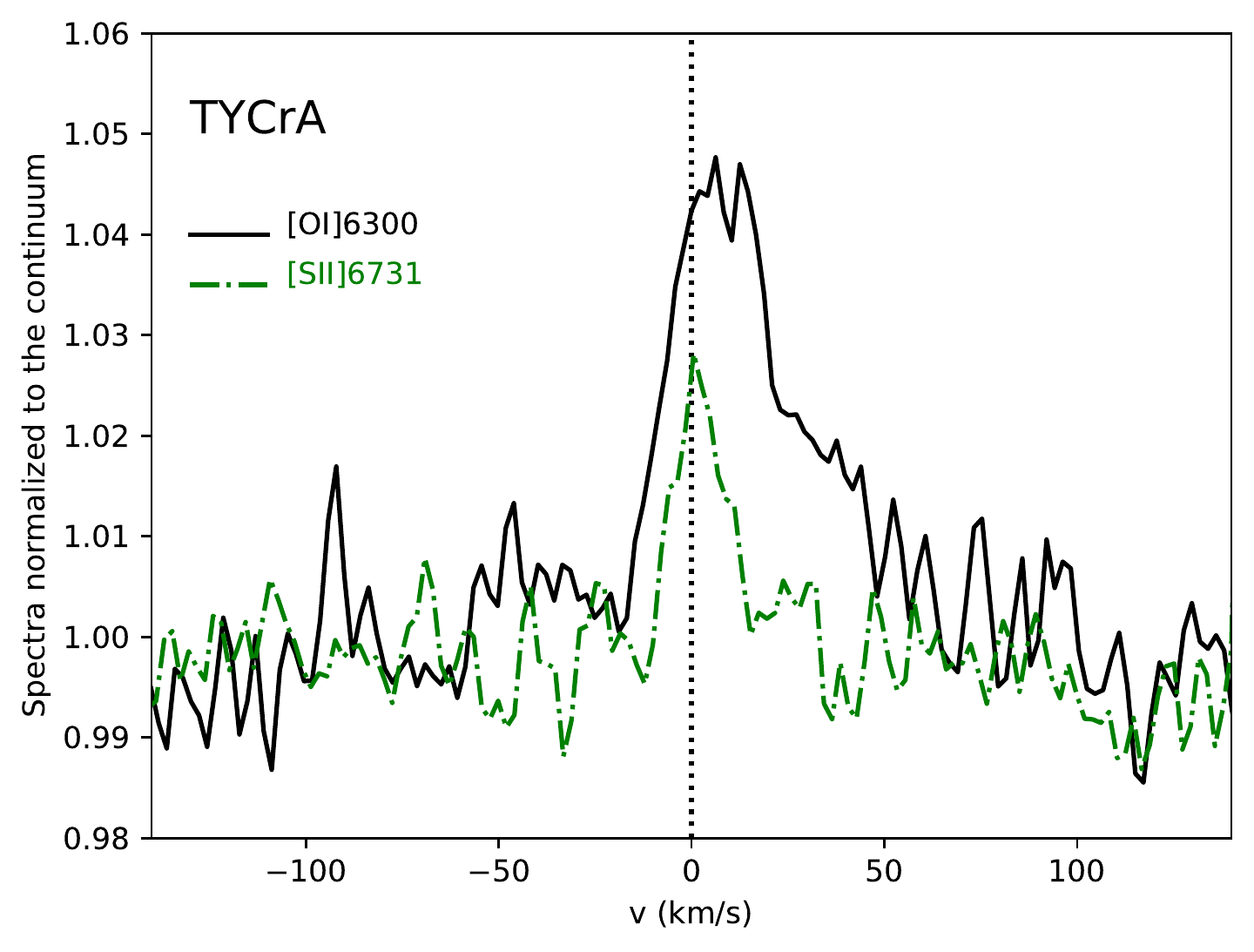}
\caption{Comparison of the \oi{} 6300\,\AA{} (black solid) and the \sii{} 6731\,\AA{} (green dot-dashed) profiles for TY~CrA. As the \sii{} 6731\,\AA{} has a critical density two orders of magnitude lower than the \oi{} 6300\,\AA{} transition, it is sensitive to jet emission. The figure illustrates that the LVC \oi{} 6300\,\AA{} is contaminated by jet emission. \label{fig:tycra}}
\end{figure}

\section{Notes on complex systems}\label{app:notes}
{\bf VW~Cha} is a triple system consisting of a K7 primary \citep{Manara2017} at 0.7$''$ from a close almost equal-mass binary  \citep[0.1$''$ separation,][]{Brandeker2001,Vogt2012}. The infrared excess is attributed to the primary \citep{Brandeker2001} which is also the component we focus on in our paper.

{\bf V4046~Sgr} is a spectroscopic ($P \sim 2.4$\, days)  binary consisting of two nearly equal-mass stars plus a possible third more distant companion \citep{Donati2011}. The stars are surrounded by a dust disk spatially resolved at millimeter wavelengths: the disk has a large inner hole (R$\sim$30\,au) and extends out to only $\sim 50$\,au \citep{Rosenfeld2013}. The gaseous  component detected in the $^{12}$CO(2-1) and $^{13}$CO(2-1) lines is much more extended (out to $\sim 370$\,au) and viewed at an inclination of $\sim 35^\circ$, consistent with that assumed to compute the central binary mass \citep{Rodriguez2010}. For the radial velocity of this system, we rely on the precise measurement from the circumbinary CO disk, a systemic LSR velocity of 2.92$\pm$0.01\,km/s which corresponds to a heliocentric velocity of -6.21\,km/s.

{\bf S~CrA} and {\bf VV~CrA} are two relatively wide binary systems with companions at 1.3$''$ and 1.9$''$ respectively \citep{Joy1945,Koresko1997,Prato2003}. Given that both systems are variable, \citet{Sullivan2019} use their 2015 infrared flux ratios to define the primary and secondary star. With this approach, S~CrA A is the NW component while VV~CrA A is the southern component. They also determine, for the first time, the spectral type for each star and find similar spectral types in the range K7-M1. As indicated in Table~\ref{tab:targets}, the Keck/HIRES spectra of S~CrA covering the \oi{} 6300\,\AA{} line are for the combined (A+B) system: both stars were positioned along the slit but the MAKEE pipeline extracted the combined flux \citep{Fang2018}. For the VV~CrA system, the optical spectrum covered only the primary southern component. In the case of the infrared survey, both binary components were placed along the slit and were extracted separately.

{\bf TY~CrA} is a quadruple system \citep[e.g.,][]{Casey1998,Chauvin2003}:
two components form a massive eclipsing double-lined spectroscopic binary  with an orbital period of  $\sim$3\,days (the primary is a B9 star of 3\,M$_\odot$ while the secondary has a mass of 1.6\,M$_\odot$); a third spectroscopic component of 1.3\,M$_\odot$ orbits around the eclipsing pair at $\sim$1\,au; a fourth more distant ($0.3''\sim 40$\,au) visual M-type companion has been also identified. The inclination of the eclipsing binary is $\sim 85^\circ$ \citep{Vanko2013} while the tertiary is highly inclined with respect to the orbit of the binary \citep{Corporon1996}.  The center of mass radial velocity for the eclipsing binary is -4.6\,km/s \citep{Casey1993}. We report this value in Table~\ref{tab:targets} and use it through our analysis.
The whole system is embedded in a bright reflection nebula, hence the flux density of TY~CrA at infrared wavelengths is rather uncertain. \citet{CurrieSicilia2011} point out that {\it Spitzer} IRAC \& MIPS flux densities at wavelengths longer than $\sim 6$\,\micron{} are unreliable due to nebular background emission, hence the lack of a spectral index in our Table~\ref{tab:targets}. \citet{Boersma2009} used a combination of ground-based infrared imaging and spectroscopy to spatially resolve the crowded environment. They infer the presence of a circumstellar disk surrounding all four stars and, thanks to a low-resolution (R$\sim$250-390) VISIR~1 spectrum, they separate the dust/continuum emission from the more extended PAH emission. It is the dust/continuum emission at 12.9\,\micron{} (Figure~5 middle panel in \citealt{Boersma2009}) which we use here to convert the \neii{} EW into a line luminosity.
TY~CrA also sports a strong X-ray emission whose properties agree with the expected combined X-ray emission of the three late-type companions \citep{ForbrichPreibisch2007}.


 {\bf T~CrA} is an F-type star that might have a companion at $>0.14''$ according to spectro-astrometry in the H$\alpha$ line  \citep{Bailey1998,Takami2003}. However, no companion has been detected using spectro-astrometry in the fundamental rovibrational band of CO
at 4.6\,\micron{} \citep{Pontoppidan2011} nor with K-band speckle imaging \citep[e.g.,][]{Koeler2008}. \citet{Wang2004} identified a Herbig-Haro knot close to T~CrA, hinting at a presence of a jet. More recently,  Whelan et al. (in prep.) discovered, through spectro-astrometry in the \oi{} and other optical forbidden lines, that the jet is nearly in the plane of the sky.  This means that the flared, massive disk surrounding T~CrA  \citep{Sicilia2013,Cazzoletti2019} is likely seen almost edge-on.

{\bf CS~Cha} is a single-line spectroscopic binary with a companion mass of at least 0.1\,M$_\odot$ and a period longer than 2482\,days \citep{Guenther2007}. Its spectral energy distribution hinted early on at the presence of a large dust cavity \citep{GauvinStrom1992} which has been recently imaged at millimeter wavelengths with ALMA \citep{Francis2020}. SPHERE polarized imagery identified an additional faint companion at a projected separation of 210\,au, outside the circumbinary disk of the primary \citet{Ginski2018}. Our high-resolution optical and infrared spectra focus on the CS~Cha spectroscopic binary and its disk.



\section{Comparison of VISIR and {\it Spitzer} \neii{} fluxes}\label{app:comparison}
Here, we compare the \neii{} 12.81\,\micron{} fluxes obtained from the {\it Spitzer}/IRS medium-resolution spectra (slit width 4.7$''$)  with those recovered with the much narrower slits of VISIR~1 ($\sim 0.4 ''$) and 2 (0.75$''$) for the targets in Table~\ref{tab:targets}. Except for TWA3A, V4046~Sgr, S~CrA, T~CrA, TY~CrA, and VV~CrA, all {\it Spitzer} \neii{} fluxes can be found in the following papers: \citet{Pascucci2007}; \citet{Guedel2010}; and \citet{Rigliaco2015}. TY~CrA was not observed with {\it Spizer}/IRS. For the other five sources we retrieve reduced {\it Spitzer} spectra from the online CASSIS database \citep{Leb2015} and, following \citet{Rigliaco2015}, we fit a Gaussian profile when the \neii{} transition is detected (for V4046~Sgr and T~CrA) or calculate a 3$\sigma$ upper limit when it is not detected (for TWA3A, S~CrA, and VV~CrA).

\begin{figure*}[h]
\plottwo{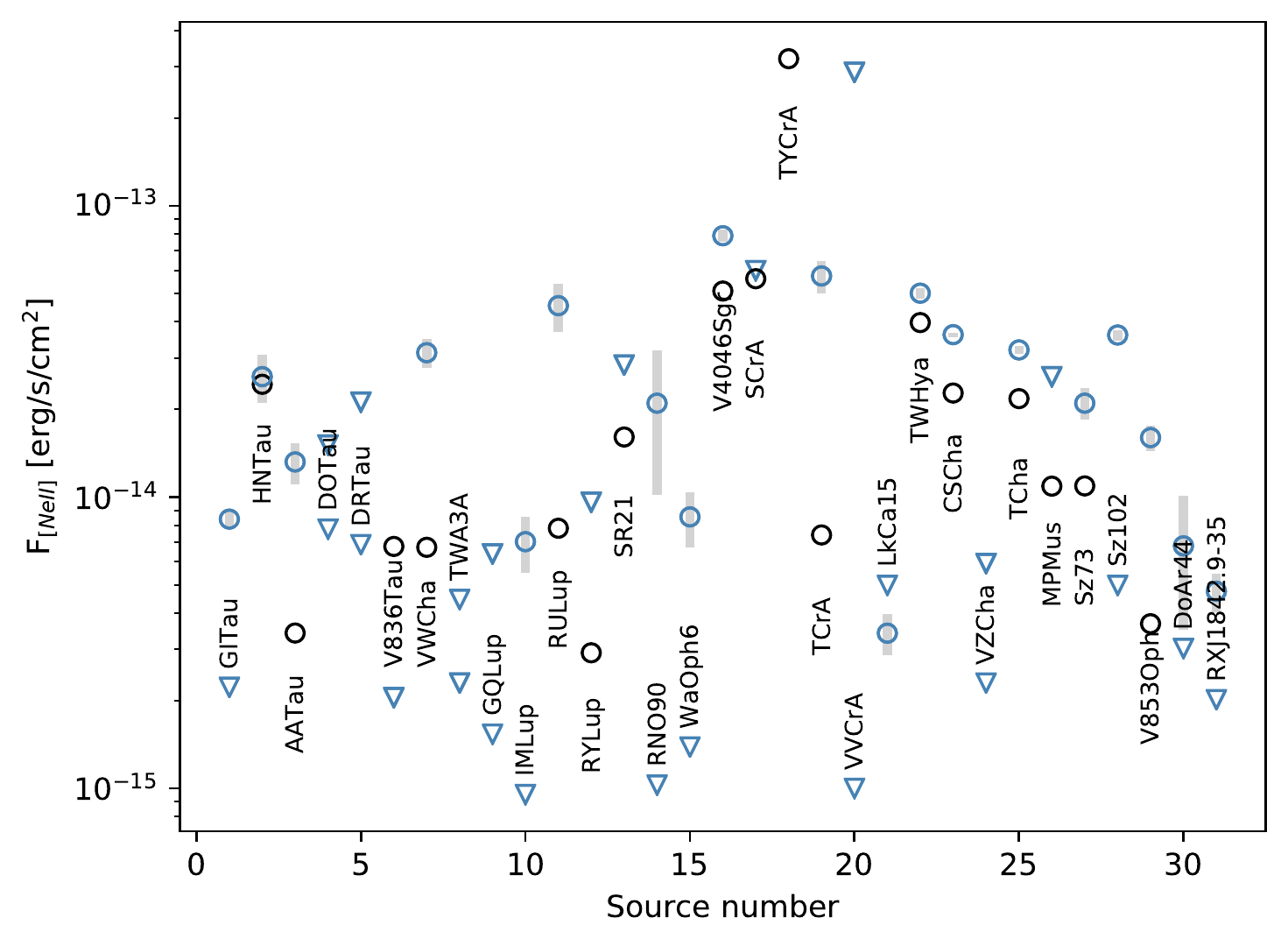}{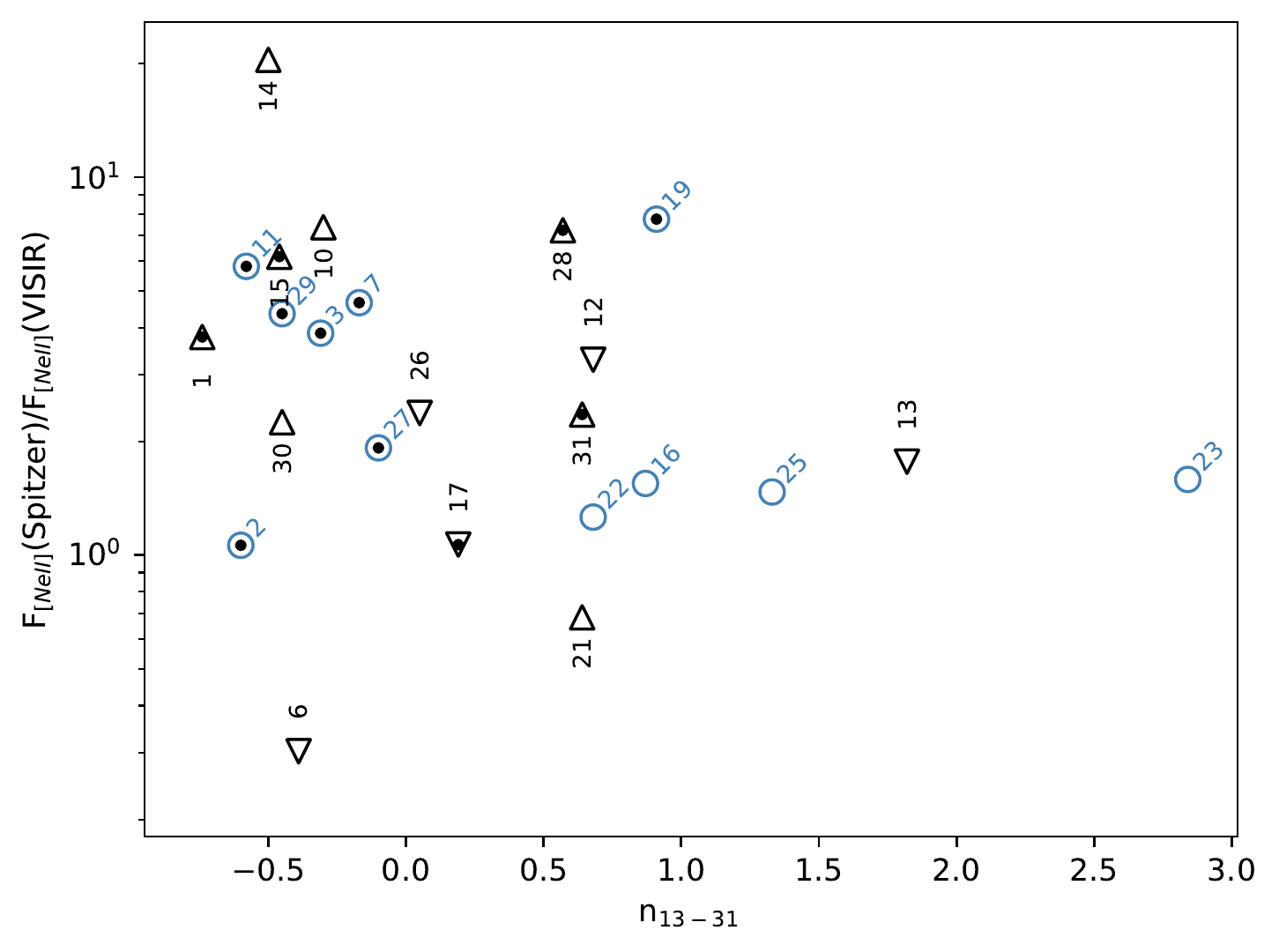}
\caption{Left panel: Comparison of {\it Spitzer}/IRS (blue) and VISIR~1 and ~2 (black) \neii{} fluxes for all sources in Table~\ref{tab:targets}. Detections are shown with circles while 3$\sigma$ upper limits with down-pointing triangles. Grey bars show the 1$\sigma$ uncertainty on the {\it Spitzer} fluxes. Right planel: \neii{} {\it Spitzer}-to-VISIR line ratios vs source infrared spectral index. Sources with detections in {\it Spitzer} and VISIR spectra are shown with circles; detections only in the {\it Spitzer} spectra with up-pointing triangles; detections only in the VISIR spectra with down-pointing triangles. A black filled circle  identifies sources with \oi{} 6300\,\AA{} HVC emission.} \label{fig:comparison}
\end{figure*}

The left panel of Figure~\ref{fig:comparison} compares the {\it Spitzer} (blue) and VISIR (black) fluxes (circles) or upper limits (downward triangles) for each source. For TY~CrA (\# 18)  we only show the VISIR detection. There are a total of 11 sources with {\it Spitzer} and VISIR \neii{} fluxes (Group~1); 8 with a {\it Spitzer} flux but no VISIR detection (Group~2: GI~Tau, IM~Lup, RNO90, WaOph6, LkCa~15, Sz~102, DoAr~44, RXJ1842.9-35);
5 with a VISIR flux but no {\it Spitzer} detection (Group~3: V836~Tau, RY~Lup, SR~21, SCr~A, MPMus); and 6 sources that do not show a \neii{} detection in the VISIR nor in the {\it Spitzer} spectrum (Group~4: DO~Tau, DR~Tau, TWA3A, GQ~Lup, VV~CrA, VZ~Cha). Except for TWA3A, the VISIR upper limits are more stringent than the {\it Spitzer} ones. Excluding Group~4 sources, we also show in the right panel of Figure~\ref{fig:comparison} the {\it Spitzer}-to-VISIR line flux ratios vs the source infrared spectral index ($n_{\rm 13-31}$): circles are used for Group~1 sources, up-pointing triangles for Group~2, and down-pointing triangles for Group~3 sources. A black filled circle identifies objects with a jet (HVC) detected in the  \oi{} 6300\,\AA{} transition, basically all Table~\ref{tab:targets} sources with $L_{\rm [OI]_{tot}} > L_{\rm [OI]_{LVC}}$. Sources with jets belong predominantly to Group~1 and 2 and, as expected, their  {\it Spitzer} flux is typically larger than the VISIR flux because of spatially extended \neii{} emission from the jet that is filtered out by the narrower slits of VISIR (see also \citealt{Sacco2012} Fig.~4 for a similar behaviour in Class~I sources). The {\it Spitzer} non detections of Group~3 sources are likely due to a combination of high infrared continuum and poor spectral resolution for the IRS which reduces the line-to-continuum ratio. Finally, sources with no jets and $n_{\rm 13-31} \gtrsim 1$ have {\it Spitzer} fluxes well within a factor of 2 of the VISIR ones. This suggests that only {\it Spitzer}/IRS sources  with significant dust depletion in their inner disk and no HVC in the \oi{} 6300\,\AA{} line should be used to further expand the VISIR sample.

With this result in mind, we cross-matched the sample of sources observed at high-spectral resolution that have only a \oi{} 6300\,\AA{} LVC detection \citep{Simon2016,Fang2018,Banzatti2019} with published and archival {\it Spitzer}/IRS spectra from sources with $n_{\rm 13-31} \gtrsim 1$. We only find 5 additional targets, all of them well known  disks with dust cavities, see Table~\ref{tab:spitzer_neii} and Figure~\ref{fig:neii_oi} in the main text.

\begin{deluxetable}{lcccc}
\tablecaption{{\it Spitzer}/IRS sources with depleted inner dust disks that can be used to expand the VISIR sample. \label{tab:spitzer_neii}}
\tablewidth{0pt}
\tablehead{
\colhead{Target} & \colhead{$n_{\rm 13-31}$} & \colhead{Log$L_{\rm [NeII]}$} & \colhead{Log $L_{\rm [OI]_{LVC}}$} &  \colhead{Ref.}
}
\startdata
CoKuTau4 & 2.27 & -5.57 & -5.55 & 1,2\\
DMTau & 1.3 & -5.45 & -5.68 & 3,4\\
GMAur & 1.8 & -5.09 & -4.55 & 3,2\\
UXTauA & 1.7 & -5.26 & -4.69 & 3,4\\
RXJ1852.33700 & 2.47 & -5.32 & -5.34 & 3,5,4\\
\enddata
\tablerefs{1. This work; 2. \citet{Simon2016}; 3. \citet{Rigliaco2015}; 4. \citet{Fang2018}; 5. \citet{Pascucci2007}}
\end{deluxetable}



\begin{thebibliography}{}
\bibitem[Acke et al.(2005)]{Acke2005} Acke, B., van den Ancker, M. E., Dullemond, C. P. 2005, A\&A, 436, 209
\bibitem[Alexander(2008)]{Alexander2008} Alexander, R. D. 2008, MNRAS, 391L, 64
\bibitem[Alexander et al.(2014)]{Alexander2014} Alexander, R., Pascucci, I., Andrews, S. et al. 2014, Protostars and Planets VI, 475
\bibitem[Alcal\'a et al.(2017)]{Alcala2017} Alcal\'a, J. M., Manara, C. F., Natta, A. et al. 2017, A\&A, 600A, 20
\bibitem[Andrews et al.(2016)]{Andrews2016} Andrews, S. M., Wilner, D. J., Zhu, Z. et al. 2016, ApJ, 820L, 40
\bibitem[Ansdell et al.(2016)]{Ansdell2016} Ansdell, M., Williams, J. P., van der Marel, N. et al. 2016, \apj, 828, 46
\bibitem[Aresu et al.(2012)]{Aresu2012} Aresu, G., Meijerink, R., Kamp, I. et al. 2012, A\&A, 547A, 69
\bibitem[Armitage(2011)]{Armitage2011} Armitage, P J. 2011, ARA\&A, 49, 195
\bibitem[Astropy Collaboration et al.(2013)]{2013A&A...558A..33A} Astropy Collaboration, Robitaille, T.~P., Tollerud, E.~J., et al.\ 2013, \aap, 558, A33
\bibitem[Bai(2016)]{Bai2016} Bai, Xue-Ning 2016, \apj , 821, 80
\bibitem[Bailer-Jones et al.(2018)]{bailer18} Bailer-Jones, C. A. L., Rybizki, J., Fouesneau, M. et al. 2018, AJ, 156, 58
\bibitem[Bailey(1998)]{Bailey1998} Bailey, J.  1998, MNRAS, 301, 161
\bibitem[Balbus \& Hawley(1991)]{BalbusHawley1991} Balbus, S. A. \& Hawley, J. F. 1991, \apj , 376, 214
\bibitem[Baldovin-Saavedra et al.(2012)]{Baldovin2012} Baldovin-Saavedra, C., Audard, M.,  Carmona, A. et al. 2012, A\&A, 543A, 30
\bibitem[Ballabio et al.(2020)]{Ballabio2020} Ballabio, G., Alexander, R. D., Clarke, C. J.  2020, MNRAS, 496, 2932
\bibitem[Banzatti et al.(2014)]{Banzatti2014} Banzatti, A., Meyer, M. R., Manara, C. F. et al. 2014, \apj , 780, 26
\bibitem[Banzatti \& Pontoppidan(2015)]{BanzattiPontoppidan2015} Banzatti, A. \& Pontoppidan, K. M. 2015, \apj , 809, 167
\bibitem[Banzatti et al.(2017)]{Banzatti2017} Banzatti, A., Pontoppidan, K. M., Salyk, C. et al. 2017, \apj , 834, 152
\bibitem[Banzatti et al.(2019)]{Banzatti2019} Banzatti, A., Pascucci, I., Edwards, S. et al. 2019, \apj , 870, 76
\bibitem[Baraffe et al.(2015)]{baraffe2015} Baraffe, I., Homeier, D., Allard, F., Chabrier, G. 2015, A\&A, 577A, 42
\bibitem[Bessell(1979)]{Bessell1979} Bessell, M. S. 1979, PASP, 91, 589
\bibitem[Blandford \& Payne(1982)]{BP82} Blandford R. D., Payne D. G., 1982, MNRAS, 199, 883
\bibitem[Boersma et al.(2009)]{Boersma2009} Boersma, C., Peeters, E., Mart\'in-Hern\'andez, N. L. et al. 2009, A\&A, 502, 175
\bibitem[Bouvier et al.(1999)]{Bouvier1999} Bouvier, J., Chelli, A., Allain, S. et al. 1999, A\&A, 349, 619
\bibitem[Brandeker et al.(2001)]{Brandeker2001} Brandeker, A., Liseau, R., Artymowicz, P., Jayawardhana, R. 2001, \apj , 561L, 199
\bibitem[Cahill et al.(2019)]{Cahill2019} Cahill, E., Whelan, E. T.; Hu\'{e}lamo, N. et al. 2019, MNRAS, 484, 4315
\bibitem[Casey et al.(1993)]{Casey1993} Casey, B. W., Mathieu, R. D., Suntzeff, N. B. et al. 1993, AJ, 105, 2276
\bibitem[Casey et al.(1998)]{Casey1998} Casey, B. W., Mathieu, R. D., Vaz, L. P. R. et al. 1998, AJ, 115, 1617
\bibitem[Cazzoletti et al.(2019)]{Cazzoletti2019} Cazzoletti, P., Manara, C. F., Baobab Liu, H. et al. 2019, A\&A, 626A, 11
\bibitem[Chauvin et al.(2003)]{Chauvin2003} Chauvin, G., Lagrange, A. -M., Beust, H. et al. 2003, A\&A, 406L, 51
\bibitem[Clarke \& Alexander(2016)]{ClarkeAlexander2016} Clarke, C. J. \& Alexander, R. D. 2016, MNRAS, 460, 3044
\bibitem[Corporon et al.(1996)]{Corporon1996} Corporon, P., Lagrange, A. M., Beust, H. 1996, A\&A, 310, 228
\bibitem[Cortes et al.(2009)]{cortes2009} Cortes, S. R., Meyer, M. R., Carpenter, J. M. et al. 2009, \apj , 697, 1305
\bibitem[Curran et al.(2011)]{Curran2011} Curran, R. L., Argiroffi, C., Sacco, G. G. et al. 2011, A\&A, 526A, 104
\bibitem[Currie \& Sicilia-Aguilar(2011)]{CurrieSicilia2011} Currie, T. \& Sicilia-Aguilar, A. 2011, \apj , 732, 24
\bibitem[de Juan Ovelar et al.(2016)]{deJuanOvelar2016}
de Juan Ovelar, M., Pinilla, P., Min, M. et al. 2016, MNRAS, 459L, 85
\bibitem[Dionatos et al.(2019)]{Dionatos2019} Dionatos, O., Woitke, P., G{\"u}del, M. et al. 2019, A\&A, 625A, 66
\bibitem[Donati et al.(2011)]{Donati2011} Donati, J. -F., Gregory, S. G., Montmerle, T. et al.  2011, MNRAS, 417, 1747
\bibitem[Dong et al.(2018)]{Dong2018} Dong, R., Najita, J. R., Brittain, S. 2018, \apj, 862, 103
\bibitem[Ercolano et al.(2008)]{Ercolano2008} Ercolano, B., Drake, J. J., Raymond, J. C., Clarke, C. C. 2008, \apj, 688, 398
\bibitem[Ercolano \& Owen(2010)]{ErcolanoOwen2010} Ercolano, B. \& Owen, J. E. 2010, MNRAS, 406, 1553
\bibitem[Ercolano \& Owen(2016)]{ErcolanoOwen2016} Ercolano, B. \& Owen, J. E. 2016, MNRAS, 460, 3472
\bibitem[Ercolano \& Pascucci(2017)]{ErcolanoPascucci2017} Ercolano, B. \& Pascucci, I. 2017, RSOS, 470114
\bibitem[Espaillat et al.(2007)]{Espaillat2007} Espaillat, C., Calvet, N., D'Alessio, P. et al. 2007, \apj , 664L, 111
\bibitem[Espaillat et al.(2011)]{Espaillat2011} Espaillat, C., Furlan, E., D'Alessio, P. et al. 2011, \apj , 728, 49
\bibitem[Espaillat et al.(2013)]{Espaillat2013} Espaillat, C., Ingleby, L., Furlan, E. 2013, \apj, 762, 62
\bibitem[Fang et al.(2018)]{Fang2018} Fang, M., Pascucci, I., Edwards, S. et al. 2018, \apj , 868, 28
\bibitem[Flaherty et al.(2017)]{Flaherty2017} Flaherty, K. M., Hughes, A. M., Rose, S. C. et al. 2017, \apj, 843, 150
\bibitem[Flaherty et al.(2020)]{Flaherty2020} Flaherty, K., Hughes, A. M., Simon, J. B. et al. 2020, in press (eprint arXiv:2004.12176)
\bibitem[Font et al.(2004)]{Font2004} Font, A. S., McCarthy, I. G., Johnstone, D., Ballantyne, D. R. 2004, \apj, 607, 890
\bibitem[Forbrich \& Preibisch(2007)]{ForbrichPreibisch2007} Forbrich, J. \& Preibisch, T. 2007, A\&A, 475, 959
\bibitem[Frank et al.(2014)]{Frank2014} Frank, A., Ray, T. P., Cabrit, S. 2014, Protostars and Planets VI, 451
\bibitem[Francis \& van der Marel(2020)]{Francis2020} Francis, L. \& van der Marel, N. 2020, \apj , 892, 111
\bibitem[Furlan et al.(2009)]{furlan2009} Furlan, E., Watson, D. M., McClure, M. K. et al. 2009, \apj , 703, 1964
\bibitem[Garcia et al.(2001)]{Garcia2001} Garcia, P. J. V., Cabrit, S., Ferreira, J., Binette, L. 2001, A\&A, 377, 609
\bibitem[Garcia Lopez et al.(2006)]{GarciaLopez2006} Garcia Lopez, R., Natta, A., Testi, L., Habart, E. 2006, A\&A, 459, 837
\bibitem[Gauvin \& Strom(1992)]{GauvinStrom1992} Gauvin, L. S. \& Strom, K. M. 1992, \apj , 385, 217
\bibitem[Ginski et al.(2018)]{Ginski2018} Ginski, C., Benisty, M., van Holstein, R. G. et al. 2018, A\&A, 616A, 79
\bibitem[Glassgold et al.(2007)]{Glassgold2007} Glassgold, A. E., Najita, J. R., Igea, J. 2007, \apj , 656, 515
\bibitem[Gorti et al.(2016)]{Gorti2016} Gorti, U., Liseau, R., S\'{a}ndor, Z., Clarke, C. 2016, SSRv, 205, 125
\bibitem[Gressel et al.(2015)]{Gressel2015} Gressel, O., Turner, N. J., Nelson, R. P., McNally, C. P. 2015, \apj , 801, 84
\bibitem[Gressel et al.(2020)]{Gressel2020} Gressel, O., Ramsey, J. P., Brinch, Ch. et al. 2020, ApJ in press (arXiv:2005.03431)
\bibitem[G{\"u}del et al.(2010)]{Guedel2010} G{\"u}del, M., Lahuis, F., Briggs, K. R. et al. 2010, A\&A, 519A, 113
\bibitem[Guenther et al.(2007)]{Guenther2007} Guenther, E. W., Esposito, M., Mundt, R. et al. 2007, A\&A, 467, 1147
\bibitem[Hartigan et al.(1987)]{Hartigan1987} Hartigan, P., Raymond, J., Hartmann, L. 1987, \apj , 316, 323
\bibitem[Hartigan et al.(1994)]{Hartigan1994} Hartigan, P., Morse, J. A., Raymond, J. 1994, \apj, 436, 125
\bibitem[Hartigan et al.(1995)]{Hartigan1995} Hartigan, P., Edwards, S., Ghandour, L. 1995, \apj , 452, 736
\bibitem[Hendler et al.(2018)]{Hendler2018} Hendler, N. P., Pinilla, P., Pascucci, I. et al. 2018, MNRAS, 475L, 62
\bibitem[Hendler et al.(2020)]{Hendler2020} Hendler, N., Pascucci, I., Pinilla, P. et al. 2020, \apj in press. (eprint arXiv:2001.02666)
\bibitem[Herczeg \& Hillenbrand(2014)]{HerczegHillenbrand2014} Herczeg, G. J. \& Hillenbrand, L. A. 2014, \apj , 786, 97
\bibitem[Hollenbach \& Gorti(2009)]{HollenbachGorti2009} Hollenbach, D. \& Gorti, U.  2009, \apj, 703, 1203
\bibitem[Hollenbach \& McKee(1979)]{HollenbachMcKee1979} Hollenbach, D. \& McKee, C. F. 1979, ApJS, 41, 555
\bibitem[Joy(1945)]{Joy1945} Joy, A. H. 1945, \apj , 102, 168
\bibitem[Kastner et al.(2010)]{Kastner2010} Kastner, J. H., Hily-Blant, P., Sacco, G. G., Forveille, T., \& Zuckerman, B. 2010, \apj, 723, L248
\bibitem[Kastner et al.(2016)]{Kastner2016} Kastner, J. H., Principe, D. A., Punzi, K. et al. 2016, AJ, 152, 3
\bibitem[Kellogg et al.(2017)]{Kellogg2017} Kellogg, K., Prato, L., Torres, G. et al. 2017, \apj, 844, 168
\bibitem[Kimmig et al. (2020)]{Kimmig2020} Kimmig, C. N., Dullemond, C. P., Kley, W. 2020, A\&A, 633A, 4
\bibitem[Koresko et al.(1997)]{Koresko1997} Koresko, C. D., Herbst, T. M., \& Leinert, C. 1997, \apj , 480, 741
\bibitem[K\'{o}sp\'{a}l et al.(2012)]{Kospal2012} K\'{o}sp\'{a}l, \'{A}., \'{A}brah\'{a}m, P.; Acosta-Pulido, J. A. et al. 2012, ApJS, 201, 11
\bibitem[K{\"o}hler et al.(2008)]{Koeler2008} K{\"o}hler, R., Neuh{\"a}user, R., Kr{\"a}mer, S., Leinert, C. et al. 2008, A\&A, 488, 997
\bibitem[Kwan \& Tademaru(1995)]{KwanTademaru1995} Kwan, J. \& Tademaru, E. 1995, \apj , 454, 382
\bibitem[Lebouteiller et al.(2015)]{Leb2015} Lebouteiller, V., Barry, D. J., Goes, C., et al. 2015, ApJS, 218, 21
\bibitem[Loomis et al.(2017)]{Loomis2017} Loomis, R. A., \"Oberg, K. I., Andrews, S. M., MacGregor, M. A. 2017, \apj , 840, 23
\bibitem[Louvet et al.(2016)]{louvet2016} Louvet, F., Dougados, C., Cabrit, S. et al. 2016, A\&A, 596A, 88
\bibitem[Mamajek et al.(2002)]{Mamajek2002} Mamajek, E. E., Meyer, M. R., Liebert, J. 2002, AJ, 124, 1670
\bibitem[Manara et al.(2017)]{Manara2017} Manara, C. F., Testi, L., Herczeg, G. J. et al. 2017, A\&A, 604A, 127
\bibitem[McClure et al.(2015)]{McClure2015} McClure, M. K., Espaillat, C., Calvet, N. et al. 2015, \apj, 799, 162
\bibitem[McGinnis et al.(2018)]{McGinnis2018} McGinnis, P., Dougados, C., Alencar, S. H. P. et al. 2018, A\&A, 620A, 87
\bibitem[Najita et al.(2009)]{najita2009} Najita, J. R., Doppmann, G. W., Bitner, M. A. et al. 2009, \apj , 697, 957
\bibitem[Najita et al.(2010)]{Najita2010} Najita, J. R., Carr, J. S., Strom, S. E. et al. 2010, \apj , 712, 274
\bibitem[Natta et al.(2014)]{Natta2014} Natta, A., Testi, L., Alcal\'{a}, J. M., Rigliaco, E., Covino, E., Stelzer, B., D'Elia, V. 2014, A\&A, 569A, 5
\bibitem[Nisini et al.(2018)]{Nisini2018} Nisini, B., Antoniucci, S., Alcal\'{a}, J. M. et al. 2018, A\&A, 609A, 87
\bibitem[Ogihara et al.(2018)]{Ogihara2018} Ogihara, M., Kokubo, E., Suzuki, T. K., Morbidelli, A. 2018, A\&A, 615A, 63
\bibitem[O'Sullivan et al.(2005)]{OSullivan2005} O'Sullivan, M., Truss, M., Walker, C., et al. 2005, MNRAS, 358, 632
\bibitem[Pascucci et al.(2007)]{Pascucci2007} Pascucci, I., Hollenbach, D., Najita, J. et al. 2007, \apj, 663, 383
\bibitem[Pascucci \& Sterzik(2009)]{PascucciSterzik2009} Pascucci, I. \& Sterzik, M. 2009, \apj , 702, 724
\bibitem[Pascucci et al.(2011)]{Pascucci2011} Pascucci, I.,  Sterzik, M., Alexander, R.~D. et al. 2011, \apj , 736, 13
\bibitem[Pascucci et al.(2014)]{Pascucci2014} Pascucci, I., Ricci, L., Gorti, U. et al. 2014, \apj, 795, 1
\bibitem[Pascucci et al.(2015)]{Pascucci2015} Pascucci, I., Edwards, S., Heyer, M. et al. 2015, \apj, 814, 14
\bibitem[Pascucci et al.(2016)]{Pascucci2016} Pascucci, I., Testi, L., Herczeg, G. J. et al. 2016, \apj , 831, 125
\bibitem[Pelletier \& Pudritz(1992)]{PelletierPudritz1992} Pelletier, G. \& Pudritz, R. E. 1992, \apj , 394, 117
\bibitem[Peterson et al.(2011)]{Peterson2011} Peterson, D. E., Caratti o Garatti, A., Bourke, T. L. et al. 2011, ApJS, 194, 43
\bibitem[Picogna et al.(2019)]{Picogna2019} Picogna, G., Ercolano, B., Owen, J. E., Weber, M. L. 2019, MNRAS, 487, 691
\bibitem[Pontoppidan et al.(2010)]{Pontoppidan2010} Pontoppidan, K. M., Salyk, C., Blake, G. A., et al. 2010, \apj , 720, 887
\bibitem[Pontoppidan et al.(2011)]{Pontoppidan2011} Pontoppidan, K. M., Blake, G. A., Smette, A. 2011, \apj , 733, 84
\bibitem[Prato et al.(2003)]{Prato2003} Prato, L., Greene, T. P., \& Simon, M. 2003, \apj , 584, 853
\bibitem[Ratzka et al.(2007)]{Ratzka2007} Ratzka, Th., Leinert, Ch., Henning, Th. et al. 2007, A\&A, 471, 173
\bibitem[Ray et al.(2007)]{Ray2007} Ray, T., Dougados, C., Bacciotti, F., Eisl\"{o}ffel, J., Chrysostomou, A. 2007 in Protostars and Planets V, 231
\bibitem[Rigliaco et al.(2013)]{Rigliaco2013} Rigliaco, E., Pascucci, I., Gorti, U. et al. 2013, \apj, 772, 60
\bibitem[Rigliaco et al.(2015)]{Rigliaco2015} Rigliaco, E., Pascucci, I., Duchene, G. et al. 2015, \apj, 801, 31
\bibitem[Rodriguez et al.(2010)]{Rodriguez2010}
Rodriguez, D. R., Kastner, J. H., Wilner, D., Qi, C. 2010, \apj , 720, 1684
\bibitem[Rosenfeld et al.(2013)]{Rosenfeld2013} Rosenfeld, K. A., Andrews, Sean M., Wilner, D. et al. 2013, \apj , 775, 136
\bibitem[Sacco et al.(2012)]{Sacco2012} Sacco, G.~G., Flaccomio, E., Pascucci, I. et al. 2012, \apj , 747, 142
\bibitem[Safier et al.(1993)]{Safier1993} Safier, P. N. 1993, \apj , 408, 115
\bibitem[Salyk et al.(2015)]{Salyk2015} Salyk, C., Lacy, J. H., Richter, M. J. et al. 2015, ApJ, 810L, 24
\bibitem[Schisano et al.(2009)]{Schisano2009} Schisano, E., Covino, E., Alcal\'{a}, J. M. et al. 2009, A\&A, 501, 1013
\bibitem[Shang et al.(2010)]{Shang2010} Shang, H., Glassgold, A. E., Lin, W.-C., Liu, C.-F. J. 2010, \apj , 714, 1733
\bibitem[Shu et al.(1994)]{Shu1994} Shu, F., Najita, J., Ostriker, E., Wilkin, F., Ruden, S., Lizano, S. 1994, \apj, 429, 781
\bibitem[Sicilia-Aguilar et al.(2013)]{Sicilia2013} Sicilia-Aguilar, A., Henning, T., Linz, H. et al. 2013, A\&A, 551A, 34
\bibitem[Simon et al.(2016)]{Simon2016} Simon, M. N., Pascucci, I., Edwards, S. et al. 2016, \apj , 831, 169
\bibitem[Simon et al.(2017)]{Simon2017} Simon, M., Guilloteau, S., Di Folco, E., et al. 2017, \apj , 844, 158
\bibitem[Simon et al.(2018)]{Simon2018} Simon, J. B., Bai, X.-N., Flaherty, K. M., Hughes, A. M. 2018, \apj , 865, 10
\bibitem[Skinner \& G{\"u}del(2017)]{SkinnerGuedel2017} Skinner, S. L. \& G{\"u}del, M. 2017, \apj , 839, 45
\bibitem[Sullivan et al.(2019)]{Sullivan2019} Sullivan, K., Prato, L., Edwards, S., Avilez, I., Schaefer, G. H. 2019, \apj , 884, 28
\bibitem[Szul\'agyi et al.(2012)]{Szulagyi2012} Szul\'agyi, J., Pascucci, I., \'Abrah\'am, P. et al. 2012, \apj , 759, 47
\bibitem[Takami et al.(2003)]{Takami2003} Takami, M., Bailey, J., Chrysostomou, A.  2003, A\&A, 397, 675
\bibitem[Teague et al.(2016)]{Teague2016} Teague, R., Guilloteau, S., Semenov, D. et al. 2016, A\&A, 592A, 49
\bibitem[Torres et al.(2006)]{Torres2006} Torres, C. A. O., Quast, G. R., da Silva, L. et al. 2006, A\&A, 460, 695
\bibitem[Tripathi et al.(2017)]{Tripathi2017} Tripathi, A., Andrews, S. M., Birnstiel, T., \& Wilner, D. J. 2017, \apj, 845, 44
\bibitem[Turner et al.(2014)]{Turner2014} Turner, N. J., Fromang, S., Gammie, C. et al. 2014, Protostars and Planets VI, 41
\bibitem[van Boekel et al.(2009)]{vanBoekel2009} van Boekel, R., G\"{u}del, M., Henning, Th. et al. 2009, A\&A, 497, 137
\bibitem[van der Marel et al.(2016)]{vanderMa2016} van der Marel, N., Verhaar, B. W., van Terwisga, S., et al. 2016, A\&A, 592, A126
\bibitem[van der Marel et al.(2018)]{vanderMa2018} van der Marel, N., Williams, J. P., Ansdell, M. et al. 2018, \apj, 854, 177
\bibitem[Va$\check{n}$ko et al.(2013)]{Vanko2013} Va$\check{n}$ko, M., Eiff, M. Ammler-von, Pribulla, T. et al. 2013, MNRAS, 431, 2230
\bibitem[Vioque et al.(2018)]{Vioque2018} Vioque, M., Oudmaijer, R. D., Baines, D., Mendigut\'{i}a, I., P\'{e}rez-Mart\'{i}nez, R. 2018, A\&A, 620A, 128
\bibitem[Vogt et al.(2012)]{Vogt2012} Vogt, N., Schmidt, T. O. B., Neuh{\"a}user, R. et al. 2012, A\&A, 546A, 63
\bibitem[Wang et al.(2004)]{Wang2004} Wang, H., Mundt, R., Henning, T., et al. 2004, \apj, 617,
1191
\bibitem[Weber et al.(2020)]{Weber2020} Weber, M. L., Ercolano, B., Picogna, G., Hartmann, L., Rodenkirch, P. J. 2020, MNRAS, 496, 223
\bibitem[Zhu et al.(2012)]{Zhu2012} Zhu, Z., Nelson, R. P., Dong, R. et al.  2012, \apj, 755, 6
\end{thebibliography}




\end{document}